 \documentclass[12pt,preprint]{aastex}

\newcommand{\gapprox}{\hbox{\lower .8ex\hbox{$\,\buildrel > \over\sim\,$}}}
\newcommand{\lapprox}{\hbox{\lower .8ex\hbox{$\,\buildrel < \over\sim\,$}}}
\newcount\countitemc
\def\itemcnumber{\number\countitemc}
\def\itemc{\global\advance\countitemc by 1   \itemcnumber}

\shorttitle{Goodness-of-Fit Tests DIFF1 and DIFF2}    % 43 letters.  ApJ allows 44. 
\shortauthors{Jeffery et al.}

\begin{document}

%% LaTeX will automatically break titles if they run longer than
%% one line. However, you may use \\ to force a line break if
%% you desire.

\title{Goodness-of-Fit Tests DIFF1 and DIFF2 for Locally-Normalized Supernova Spectra}

\author{
David J.~Jeffery\altaffilmark{1},
Wesley Ketchum\altaffilmark{1},
David Branch\altaffilmark{1},
E. Baron\altaffilmark{1},
Abouazza Elmhamdi\altaffilmark{2,3},
\&
I.J.~Danziger\altaffilmark{3}
}

%% Use \author, \affil, and the \and command to format
%% author and affiliation information.
%% Note that \email has replaced the old \authoremail command
%% from AASTeX v4.0. You can use \email to mark an email address
%% anywhere in the paper, not just in the front matter.
%% As in the title, you can use \\ to force line breaks.

%% Notice that each of these authors has alternate affiliations, which
%% are identified by the \altaffilmark after each name.  Specify alternate
%% affiliation information with \altaffiltext, with one command per each
%% affiliation.

\altaffiltext{1}{Homer L. Dodge Department of Physics \& Astronomy, University of Oklahoma,
                440 W. Brooks St., Norman, Oklahoma 73019, U.S.A.}
\altaffiltext{2}{ICTP- The International Centre for theoretical Physics, Strada Costiera 11 34014, Trieste, Italy}
\altaffiltext{3}{INAF-Osservatorio Astronomico di Trieste, via G.B.~Tiepolo 11, 34131 Trieste, Italy}
%\altaffiltext{3}{Patron, Spring Mountain Pub \& Brewery}

\begin{abstract}
    Two quantitative tests DIFF1 and DIFF2 for
measuring goodness-of-fit between two locally-normalized supernova spectra are presented. 
Locally-normalized spectra are obtained by dividing a spectrum by the same spectrum
smoothed over a wavelength interval relatively large compared to line features, but 
relatively small compared to continuum features.
DIFF1 essentially measures the mean relative difference between the line patterns of
locally-normalized spectra and DIFF2 is DIFF1 minimized by a relative logarithmic 
wavelength shift between the spectra:  the shift is effectively an artificial relative Doppler shift. 
Both DIFF1 and DIFF2 measure the physical similarity of line formation, and thus of supernovae.
DIFF1 puts more weight on overall physical similarity of the supernovae than DIFF2  
because the DIFF2 shift compensates somewhat for some physical distinction in the supernovae.
Both tests are useful in ordering supernovae into empirical groupings for further analysis. 
We present some examples of locally-normalized spectra for Type~IIb supernova SN~1993J
with some analysis of these spectra.
The UV parts of two of the SN~1993J spectra are {\it HST} spectra that have not been published before.
We also give an example of fitted locally-normalized spectra
and, as an example of the utility of DIFF1 and DIFF2, some preliminary statistical
results for hydrogen-deficient core-collapse (HDCC) supernova spectra.
This paper makes use of and refers to material to found at the first author's online supernova spectrum database
SUSPEND (SUpernovae Spectra PENDing further
analysis).\footnote{See http://www.nhn.ou.edu/{\tt\~{}}jeffery/astro/sne/spectra/spectra.html\ .}

\end{abstract}

%% Keywords should appear after the \end{abstract} command. The uncommented
%% example has been keyed in ApJ style. See the instructions to authors
%% for the journal to which you are submitting your paper to determine
%% what keyword punctuation is appropriate.


\section{INTRODUCTION\label{introduction}}

     An ordinary $\chi^{2}$-like test measuring the goodness-of-fit
of two supernova spectra often
fails to be consistent with what the eye sees qualitatively:   good agreement to
the eye can be poor or moderate agreement by a $\chi^{2}$-like test. 
Since the human eye is an excellent pattern recognization tool, judgments based
on eye comparisons informed by a specialist's understanding of spectrum formation
are often preferred over $\chi^{2}$-like tests or other formulaic  
%% defn:  characterized by or in accordance with some formula
tests in identifying similarity and/or goodness-of-fit between two spectra.
Human judgment, of course, is qualitative and to some degree subjective.
Thus, it would be advantageous to have tests that measure spectrum similarity
more consistently with what the eye sees and yet be quantitative and objective.
Such tests could be applied in an automated fashion and, one hopes, would show correlations
missed by the eye.

     In this paper, we present two quantitative tests DIFF1 and DIFF2 for
measuring goodness-of-fit between two supernova spectra.
The tests are both measures of the relative difference in line patterns.
(The formulae for DIFF1 and DIFF2 are given, respectively, in \S\S~4.1 and~4.2.)
Both tests depend on what we call local normalization which largely removes continuum features
without distorting the line features too much.
Local normalization reduces the continuum shape to an apparent flat line of height 1 by dividing the
original spectrum by a version of the spectrum smoothed over 
wavelength interval relatively large compared to line features, but
relatively small compared to continuum features.
Ordinary (or global) normalization is just to divide the spectrum by a constant chosen for some purpose.
The near-equivalent of local normalization has frequently been used for synthetic spectrum
calculations simply by inputting a flat continuum at the base of a model atmosphere or
by dividing a synthetic spectrum by a known continuum.
In the context of synthetic supernova spectra see, e.g., \citet{dessart2005} for this kind of procedure.
But note that local normalization is a particular procedure applied to spectra with line
features and it cannot be exactly equivalent to using a flat continuum or dividing a
spectrum by a known continuum in synthetic spectrum calculations. 

     DIFF1 and DIFF2 evolved from the DIFF test presented by \citet{branch2006b}.
DIFF1 essentially measures the mean relative difference between the line patterns of
locally-normalized spectra and DIFF2 is DIFF1 minimized by a relative logarithmic
wavelength shift between the spectra:  the shift is effectively an artificial relative Doppler shift.
From this description, the reader can roughly understand both tests before knowing the test formulae.
Both DIFF1 and DIFF2 measure the physical similarity of line formation, and thus of supernovae.
DIFF1 puts more weight on overall physical similarity of the supernovae than DIFF2
because the DIFF2 shift compensates somewhat for some physical distinction in the supernovae.
Both tests are useful in ordering supernovae into empirical groupings for further 
analysis.\footnote{While this paper was in preparation, analysis tools for supernova
spectra that are in some respects similar to DIFF1 and DIFF2 with local normalization
were described by St\'ephane Blondin \citep{blondin2007b} and Avet Harutyunyan 
\citep{harutyunyan2005} at the conference {\it The Multicoloured Landscape of Compact
Objects and their Explosive Origins}, 
Cefal\`u, Sicily, 2006 June 11--24, URL:  http://www.mporzio.astro.it/cefalu2006/ .} 

     There are several reasons for local normalization (or some similar continuum elimination
technique) in comparative supernova spectrum analysis.
First, the intrinsic supernova continuum can often be quite uncertain.
Often the main reason for this is uncertainty in the reddening correction.
Error in reddening correction can affect, for example, the $B-V$ color
and other colors (particularly toward the blue) by a significant fraction of a magnitude or more.
Uncertainty in reddening is particularly a problem for core-collapse supernovae (i.e.,
Types II-P, II-L, IIn, IIb, Ib, Ic, and Ic hypernovae) which tend to arise in or near star-forming regions in their
host galaxies.
Foreground Galactic reddening can be reasonably confidently corrected for using the results of
\citet{schlegel1998} for Galactic reddening and \citet{cardelli1989} and \citet{o'donnell1994} for the reddening law,
but the host galaxy reddening can usually only be corrected
for with great uncertainty relying on interstellar lines in the supernova spectra or, with less uncertainty
depending on cases, on spectral modeling.
For example, early spectral modeling of Type~II-P supernovae can constrain reddening
\citep{baron2003,baron2004,baron2005}.   
Nevertheless, in spectral modeling one would much prefer
to have reddening as a given rather than as a parameter to be fitted for---or not to be a consideration
which is what local normalization helps toward. 
Core-collapse supernova types are all rather heterogeneous types, and so judging reddening
by a comparison of a supernova to other examples of its type is not usually conclusive. 
Type~Ia supernovae (SNe~Ia) are usually much less reddened as they do not preferentially
arise in or near star forming regions and the homogeneity of most of them \citep[e.g.,][]{branch2006b}
allows reddening correction by comparison in some cases.
But even SNe~Ia can be highly reddened (e.g., \object{SN 1986G} \citep{phillips1987}) 
and there are peculiar SNe~Ia 
\citep[e.g.,][]{branch2006b} for which comparisons to determine reddening can be uncertain. 
Thus, the reddening correction for SNe~Ia is often uncertain.

    In addition to reddening, the continuum can be in error because of errors in the broad-band calibration 
of the spectrum.
A particular problem in achieving good calibration arises because supernovae are transient, 
time-dependent sources:
this means one usually cannot simply replace a bad observation by
a good one for a given epoch and this limits the ability to achieve good calibration.
Another related particular problem arises from the fact that bright well-observed supernovae 
are relatively rare.
Thus, many well-observed supernovae are from earlier instrumentation epochs.
One would like to use the data from these supernovae for analysis since they are to some degree unique even
though the data may have very uncertain calibration by modern standards.
% One cannot simply replace the older sets by newer sets since no supernova ever exactly repeated.

    Reddening and calibration errors tend to affect the spectrum continuum multiplicatively:
i.e., they change flux level by a scaling factor that varies slowly with wavelength. 
These kind of uncertainties are ideally, and nearly practically, irrelevant to the process of
local normalization (see \S~2.1), and thus cannot much affect analysis with the locally-normalized spectra.

     A kind of continuum uncertainty that is not eliminated by local normalization is
the additive uncertainty due to contamination by extraneous sources.
(See \S~2.1 for why this is so.)
The most common contamination is from the host galaxy emission.
When a supernova is at its brightest this contamination is often small, but in later phases
when the supernova dims it can become significant and dominant.
Host galaxy contamination can be subtracted off using spectra of the galaxy when the supernovae
is absent, but it is not always clear to analyzers of supernova spectra when this has been done.
Another kind of contamination is peculiar to supernovae associated with gamma-ray bursts (GRBs):
such supernovae may all be what are now called hypernovae (since they are very energetic and perhaps
very massive) or as we prefer Type~Ic~hypernovae since
they seem to lack conspicuous lines of hydrogen and helium and the strong Si~II~$\lambda6355$ line.
(Lack of conspicuous hydrogen and helium lines together with the lack of a strong Si~II~$\lambda6355$ line
are the defining observational characteristics of Type~Ic supernovae.)
The contamination is the UVOIR (ultraviolet-optical-infrared) afterglow of the GRB itself which
tends to be a line-free continuum. 
Correction for this kind of contamination can be done, but with some uncertainty \citep{matheson2003}.
% on matheson2003 p. 403, a power-law afterglow continuum is used:  justifies ``line-free''.
Yet another contamination (which unfortunately also affects the line pattern) is light echoing 
caused by supernova
light from earlier phases reflected off dust clouds and added to the light of the current supernova phase.
Light echoes, if recognized, can be corrected for, but with some error, of course.
For the developments in this paper, we assume that contamination of all kinds can be adequately corrected for.
 
     Because of all the effects mentioned above, the continuum shape of supernova spectra may be incorrect and
misleading about the degree of similarity between spectrum pairs. 
On the other hand the line pattern in the spectra should be better for determining similarities since
this is less affected by broad-band uncertainties in reddening and calibration.  %, and contamination.
For example, if a spectrum pair had an identical line pattern, but different continuum shapes,
% or level,
one would have good reason to believe that the supernovae are, in fact, 
highly similar and at a similar phase in their
evolution and that the discrepancy in continuum shape was just caused by reddening correction and/or
calibration error. % and in continuum level by contamination error. 
In the contrasting case, where the line patterns are very different, but the continuum shapes  % or levels
are identical, one would conclude the supernovae are not very similar or at different phases in
their evolution despite the similar continua.
% at very early hot times the optical can be line-free and later still very hot have lines but the
% same Rayleigh-Jeans continuum.  It's possible in principle.
In both extreme cases, it is the line patterns that provide the decisive evidence, not the continua.
Thus, if one can adequately correct or neglect contamination, analysis with locally-normalized spectra
should give good insight into the intrinsic similarities and differences among supernovae.
It is a basic premise of this paper that line pattern is a much better signature
of intrinsic supernova behavior than continuum shape, and so eliminating
continuum shape information is not too important for spectrum comparisons.

     A second reason for using local normalization applies even when the continuum
shapes are assumed to be well known.
If one is searching just for similarity of line patterns as a clue to physical similarities,
then varying continuum shapes tend to obscure the similarities of the patterns both to the
eye and formulaic tests.
Being able to measure similarity of line patterns for heterogeneous samples of supernovae
from varying phases is important in studying the time-varying supernova structure. 
Especially for core-collapse supernovae in which even members of the same type show
considerably individuality and in which the time coverage of their evolution can often
be very incomplete, local normalization could become an important tool in analysis. 

     A third reason, for local normalization is that spectra are frequently analyzed using
synthetic spectra calculated from highly simplified radiative transfer:  e.g., analyses done
using the parameterized code SYNOW 
\citep[e.g.,][and references therein]{branch2003,branch2005} 
Such simplified radiative transfer does not treat continuum radiative transfer with high physical realism
and in fits of synthetic to observed spectra mismatches in the continuum in some regions are
obvious and are not considered very important in the important results derived from the synthetic
spectrum analysis:  the important results being line identification and ejecta velocity structure. 
Local normalization of both observed and synthetic spectra can be used to reduce the
distraction of mismatching continuum shapes \citep[e.g.,][]{parrent2007}. 

     In \S~2 of this paper, we discuss the theoretical basis for local normalization
and how we actually carry out local normalization.
Section~3 gives examples of locally-normalized spectra for the Type~IIb supernova \object{SN 1993J}
with some analysis of these spectra.
The UV parts of two of the SN~1993J spectra are {\it Hubble Space Telescope} ({\it HST}$\,$) 
spectra that have not been published before.
Also in \S~3 (\S~3.5) is an investigation of the continuum independence of local normalization.
The DIFF1 and DIFF2 formulae are presented in \S~4, where we also present an example of fitted locally-normalized spectra.
As an example of the utility of DIFF1/2, \S~5 gives some preliminary statistical results for the spectra
of supernovae of Types~IIb, Ib, Ic, and Ic~hypernovae.
We will collectively refer to these supernovae as hydrogen-deficient core-collapse (HDCC) supernovae.
Conclusions and discussion are given in \S~6.

     This paper makes use of and refers to material found at the first author's online supernova spectrum database   
SUSPEND (SUpernovae Spectra PENDing further analysis).
The URL for SUSPEND is given in a footnote to the abstract. 

\section{LOCAL NORMALIZATION}

    The basic procedure of local normalization is to divide a spectrum by a smoothed version of itself
where the smoothing length is sufficiently large that line features are largely erased, but the continuum shape
is largely unaffected, in the smoothed version.
We call the smoothing for local-normalization
large-scale smoothing to distinguish it from the small-scale smoothing
used to suppress noise.
The spectrum resulting from the division has ideally a continuum level of 1 everywhere as judged by
the eye. 

    In \S~2.1, we discuss the theoretical basis for local normalization.
In \S~2.2, we discuss our actual procedure for carrying out local normalization.

\subsection{The Theoretical Basis for Local Normalization}

    For concreteness, in our discussion let us model an observed spectrum $f_{\lambda,\rm obs}$ by the following 
heuristic formula:
\begin{equation}
f_{\lambda,\rm obs}=S_{\lambda}f_{\lambda,\rm con}f_{\lambda,\rm lin}+C_{\lambda} \,\, ,
\label{eq-f-obs}
\end{equation}
where $S_{\lambda}$ is some kind of multiplicative scaling error which could be the effect of
unknown reddening or calibration error, $f_{\lambda,\rm con}$ is the intrinsic continuum spectrum,
$f_{\lambda,\rm lin}$ is the line spectrum,
$C_{\lambda}$ is some extraneous and unknown contamination flux, and
\begin{equation}
f_{\lambda,\rm int}=f_{\lambda,\rm con}f_{\lambda,\rm lin} \,\, ,
\end{equation}
is the intrinsic spectrum.
The line spectrum is defined by saying it has a height of 1 when it is smoothed 
over some specified smoothing length that is of order the size of the full width
of an individual line profile:  thus
\begin{equation}
\langle f_{\lambda,\rm lin} \rangle = 1\,\, ,
\label{eq-line-spectrum-smoothed}
\end{equation}
where the angle brackets indicate smoothing with the required smoothing length.
We make the assumption that $S_{\lambda}$,  $f_{\lambda,\rm con}$, and $C_{\lambda}$ are all
sufficiently slowly varying with wavelength that smoothing them over
the smoothing length leaves them effectively unchanged. 
(We are neglecting the possibility of uncorrected-for light-echo contamination (see \S~1).) 
Now say we smooth the observed spectrum over the smoothing length.
We obtain
\begin{equation}
\langle f_{\lambda,\rm obs}\rangle=\langle S_{\lambda}f_{\lambda,\rm con}f_{\lambda,\rm lin}\rangle +C_{\lambda} \,\, ,
\label{eq-f-obs-smoothed}
\end{equation} 
where we have used the assumption that $C_{\lambda}$ is unaffected by smoothing.
% We assume no correlation between $S_{\lambda}f_{\lambda,\rm con}$ and $f_{\lambda,\rm lin}$.
%% The above remark doesn't add anything to the cogency. 
Assuming that $S_{\lambda}f_{\lambda,\rm con}$ can be approximated as constant over line widths,
we now obtain 
\begin{equation}
\langle S_{\lambda}f_{\lambda,\rm con}f_{\lambda,\rm lin}\rangle 
= \langle S_{\lambda}f_{\lambda,\rm con}\rangle\langle f_{\lambda,\rm lin}\rangle 
= S_{\lambda}f_{\lambda,\rm con}   \,\, ,
\label{eq-f-obs-smoothed-b}
\end{equation} 
where we have used the assumptions that $S_{\lambda}$ and $f_{\lambda,\rm con}$ are unaffected by smoothing
and equation~(\ref{eq-line-spectrum-smoothed}).
Making use of equations~(\ref{eq-f-obs}), (\ref{eq-f-obs-smoothed}), and~(\ref{eq-f-obs-smoothed-b}), and
our concept of local normalization,
the locally-normalized version of $f_{\lambda,\rm obs}$ (which we denote by $f_{\lambda,\rm LN}$) is given by
\begin{equation}
f_{\lambda,\rm LN}
=
{f_{\lambda,\rm obs}\over \langle f_{\lambda,\rm obs}\rangle}
= { S_{\lambda}f_{\lambda,\rm con}f_{\lambda,\rm lin}+C_{\lambda}
   \over
    S_{\lambda}f_{\lambda,\rm con}+C_{\lambda} }  
=f_{\lambda,\rm lin}
  \left[1+C_{\lambda}/(S_{\lambda}f_{\lambda,\rm con}f_{\lambda,\rm lin})\over
        1+C_{\lambda}/(S_{\lambda}f_{\lambda,\rm con})                    \right] \,\, .
\label{eq-locally-normalized-line-spectra}
\end{equation}
 
       What we would really like to have is the line spectrum $f_{\lambda,\rm lin}$
(as we have defined it above) and what we can obtain from measurements is $f_{\lambda,\rm LN}$.
The line spectrum $f_{\lambda,\rm lin}$ is what we believe contains much of the information about the object 
that we are interested in.  
The $f_{\lambda,\rm lin}$ and $f_{\lambda,\rm LN}$ spectra
are equal if contamination is zero no matter what $S_{\lambda}$ and $f_{\lambda,\rm con}$ may be given
our assumptions.
(The equality is only exact in the spectrum model of this discussion.
In reality, it can only be approximate since line and continuum behavior cannot be completely
separated.)
Thus, the effect of any scaling errors are eliminated from the locally-normalized spectra
in the absence of contamination.
This is one of the great benefits of using locally-normalized spectra in spectrum analysis. 

       If $C_{\lambda}/(S_{\lambda}f_{\lambda,\rm con})<<1$ and 
$f_{\lambda,\rm lin}$ not much smaller than 1, then
\begin{equation}
f_{\lambda,\rm LN}
\approx f_{\lambda,\rm lin}
  \left[1+{C_{\lambda}\over S_{\lambda}f_{\lambda,\rm con}}\left({1\over f_{\lambda,\rm lin}}-1\right)\right] \,\, , 
\label{eq-locally-normalized-line-spectra-1st}
\end{equation}
where we have expanded equation~(\ref{eq-locally-normalized-line-spectra}) 
to first order in $C_{\lambda}/(S_{\lambda}f_{\lambda,\rm con})$.   % See Arfken-238.
We see that contamination error only cancels to zeroth order.
Nevertheless, if $C_{\lambda}/(S_{\lambda}f_{\lambda,\rm con})$ is sufficiently small, then
contamination will have only a small effect on the locally-normalized spectrum.
On the other hand, if $C_{\lambda}/(S_{\lambda}f_{\lambda,\rm con})$ becomes large compared to 1
and $f_{\lambda,\rm lin}$,
the locally-normalized spectrum tends to have value 1 everywhere and be almost independent of the line spectrum
as one can see from equation~(\ref{eq-locally-normalized-line-spectra}).

     In the rest of this paper, we assume contamination is, in fact, negligible or
has been adequately corrected for, and thus that $C_{\lambda}$ is approximately zero.
Given this assumption, the locally-normalized spectra are approximately
the line spectra as we have defined them above.
 
\subsection{The Procedure of Carrying Out Local Normalization}

    There is probably no absolute optimum way of carrying out the basic procedure of local
normalization.
We have developed a prescription, however, that seems very adequate.
The large-scale smoothing is box-car-like where we march through the spectrum wavelength by
wavelength and make use of a box-car interval about each wavelength
to construct a smoothed flux value at that wavelength. 
The size of the box-car interval is the smoothing length.

     The choice of the smoothing length is a key point.
If one makes the smoothing length too small, line features tend to get suppressed 
in the locally-normalized spectra:  in the limit that the smoothing length goes to zero, the locally-normalized
spectra become 1 everywhere.
On the other hand, if one makes the smoothing length too large, then the locally-normalized spectra
retain continuum features:  in the limit that the smoothing length goes to infinity, the locally-normalized
spectra are just ordinary (or globally) normalized spectra.
Unfortunately in supernova spectra, the wavelength scale of the P~Cygni and emission
line profiles can be very large because of the large Doppler shift velocities in the supernova ejecta.
(Supernova lines in the photospheric or optically-thick phase are broad P-Cygni lines with
a blueshifted absorption and an emission feature centered roughly on the rest-frame line-center wavelength.
See, e.g., \citet[p.~173--194]{jeffery1990} for a discussion of P-Cygni line formation in supernovae.)
Usually, the largest velocities one needs to consider are of order $30,000\,{\rm km\,s^{-1}}$ 
%  (reference???)   ! I guess the collective expertise of the authors is authority enough.
and these lead to relative Doppler shifts of line opacity wavelength of $v/c\sim 0.1$ or of order $10\,$\%.
However, line features can form at velocities of up to perhaps $50,000\,{\rm km\,s^{-1}}$
in Type~Ic hypernovae \citep[e.g.,][]{mazzali2000} at early times and of up to at least 
$40,000\,{\rm km\,s^{-1}}$ in other kinds of supernovae if seen at sufficiently
early times in the UV as evidenced by the P~Cygni line of the
resonance multiplet Mg~II $\lambda2797.9$ (e.g., \citealt[p.~30]{wiese1969}; \citealt[p.~10]{moore1968}) 
seen in the earliest {\it IUE} spectrum of Type~II supernova
\object{SN 1987A} \citep[e.g.,][]{pun1995}.
With Doppler shift velocity of $50,000\,{\rm km\,s^{-1}}$, a relative wavelength shift would be 
$v/c\sim 1/6$ or of order $17\,$\%.
Since one has to consider opacity shifted both to the blue and the red, the relative wavelength 
width of a line profile in extreme cases could be of order $34\,$\%.
Continuum features that vary on the scale of $34\,$\% of wavelength certainly exist.
Thus, for supernovae at least in extreme cases, one cannot completely decouple line
and continuum behavior.
 
     One must make a choice for the smoothing length for locally-normalized spectra
that does not suppress line behavior too much
and that does not let too much continuum behavior leak in.
There is probably no absolute optimum choice for all cases. 
We have found that a relative smoothing length of $30\,$\% of the current wavelength in the smoothing
procedure is good:  the smoothing region extends $15\,$\% blueward and $15\,$\% redward of the current wavelength
as one marches through the wavelength points.
The choice of $30\,$\% is motivated as follows.
The continuum level determined by blackbody-behavior and reddening \citep{cardelli1989,o'donnell1994}
is relatively constant over a region of $30\,$\% surrounding a wavelength.
On the other hand, as mentioned above,
usually the largest velocities one needs to consider are of order $30,000\,{\rm km\,s^{-1}}$ 
%  (reference???)   ! I guess the collective expertise of the authors is authority enough.
and these lead to relative Doppler shifts of line opacity wavelength of $v/c\sim 0.1$ or of order $10\,$\%:
the whole line profile in this case will vary over $\sim 20\,$\% of the current wavelength.
Thus a relative smoothing length of $30\,$\% should be effective at smoothing away most line features
in the smoothed spectrum, but not smoothing the continuum shape too much.

     To calculate the large-scale smoothed spectrum itself, a common approach is to simply numerically integrate
(e.g., using the trapezoid rule)
the spectrum over the box-car interval centered on the current wavelength and divide by the box-car interval
to obtain an average flux for the box-car interval which one assigns to the current wavelength.
For ordinary small-scale smoothing, nothing special needs to be done about the ends of the spectrum even though
they can be treated somewhat wrongly.
As an end is approached, the box-car interval gets progressively truncated at that end and the calculated average flux 
is progressively a less good approximation for the smoothed flux at the current wavelength:
it is the smoothed flux for some point in the box-car interval farther from the end.
In small-scale smoothing, the smoothing length is comparatively short and somewhat bad behavior at the 
spectrum ends is usually unimportant.
But with a large smoothing length, such as we require for local normalization, this bad behavior at the 
ends can significantly degrade local normalization near the ends.
If the spectrum continuum is relatively flat near the ends, there may be no noticeable degradation.
But if the continuum is rising/falling at an end, then the smoothed spectrum becomes too small/large at that end and
the locally-normalized spectrum continuum can appear to be rising/falling from 1 at the end.  
To avoid this problem, we fit a line using least-squares to the spectrum in the box-car interval and determine the
smoothed spectrum at the current wavelength from this fitted line.
(Actually we fit a line to logarithmic flux as a function of logarithmic wavelength.  
This means we are fitting the flux to a power-law function of wavelength.
See below.)
In the interior of the spectrum, this least-squares line fitting
gives almost the same result as calculating an average flux for the box-car interval,
but at the spectrum ends it usually prevents the locally-normalized continuum from departing obviously from 1. 
Since spectra are not necessarily very linear (or like a power-law function) 
over the box-car interval, it is not immediately obvious that this fitting 
of a line (actually a power-law function in our
calculations) should always work well.
But the resulting locally-normalized spectra seem generally pleasing to the eye 
including at the ends and we accept the eye's judgment for our prescription. 
 
     In plotting spectra, we often prefer to plot logarithmic flux versus logarithmic wavelength.
The choice of logarithmic wavelength owes to the fact that line profile widths are determined by
large Doppler shifts of line opacity wavelength from the rest-frame line opacity wavelength.
For example, say a particular component of the line profiles forms in the ejecta moving 
at velocity $v$ along the line-of-sight
counting velocity as positive in the direction away from the observer.
For a given line of rest-frame line-center wavelength $\lambda_{\rm line}$, the component will appear in a spectrum at
\begin{equation}
\lambda=\lambda_{\rm line}\sqrt{1+v/c \over 1-v/c} \approx \lambda_{\rm line} (1+v/c) \,\, , 
\label{eq-Doppler-formula}
\end{equation}
where we have used the relativistic Doppler shift formula
(e.g., \citealt{lawden1975}, p.~78;  \citealt{mihalas1978}, p.~495) and
the 1st order (in $v/c$) Doppler shift formula.
(The 1st order Doppler shift formula is usually adequate for supernovae.)
From equation~(\ref{eq-Doppler-formula}), we see for a set of lines that 
the components in a spectrum plot against wavelength will appear at different 
wavelength shifts from the line-center wavelengths.
Thus, the line profiles will tend to have varying widths that scale with $\lambda_{\rm line}$.
In spectrum figures,  %  plotted using wavelength,   % redundant with 2 sentences above.
this has an unpleasing effect of making the blue side of a spectrum look 
relatively crowded with line features and the red side look relatively uncrowded with line features. 
If one plots using logarithmic wavelength, then the aforementioned particular component is found at 
\begin{eqnarray}
\log(\lambda)
&=&\log\left(\lambda_{\rm line}\sqrt{1+v/c \over 1-v/c}\,\right)
=\log(\lambda_{\rm line})+\log\left(\sqrt{1+v/c \over 1-v/c}\,\right) \cr
&\approx&\log[\lambda_{\rm line}(1+v/c)]
=\log(\lambda_{\rm line})+\log(1+v/c)
\approx \log(\lambda_{\rm line}) + \left({v\over c}\right)\log(e)  \,\, ,
\end{eqnarray}
where we have made an expansion of the logarithm to 1st order in $v/c$ which is consistent in order of expansion
with the 1st order Doppler shift formula.
We see that in logarithmic wavelength, the shift of the particular component is the same for all line profiles
(from relativistic Doppler shift formula) 
and that this shift is approximately proportional to $v/c$.
Thus, the line profiles will tend to have the same width across the spectrum when plotted
using logarithmic wavelength.
Varying line strength and line blending, of course, prevent the profiles from having an identical appearance
when plotted versus logarithmic wavelength.
Using logarithmic wavelength gives the plotted spectrum a balanced and pleasing appearance.
The above argument for plotting spectra versus logarithmic wavelength applies for both 
ordinary spectra and locally-normalized spectra.

     In plotting ordinary spectra (i.e., not locally-normalized spectra), 
we often make the choice of plotting logarithmic flux since 
it gives a more equal appearance to small and large features of a spectrum.
Say we take equation~(2) (see \S~2.1) for the intrinsic flux (which for the sake of argument we assume that we know)
and take the logarithm:  we get
\begin{equation}
\log\left(f_{\lambda,\rm int}\right)=\log\left(f_{\lambda,\rm con}\right)+\log\left(f_{\lambda,\rm lin}\right) \,\, . 
\end{equation}
From equation~(10), we see that line profiles of the same relative size (i.e., the same $f_{\lambda,\rm lin}$ profile)
will have the same absolute size when logarithmic flux is plotted since the line profiles are just
added to the continuum for logarithmic flux and not multiplied by the continuum as in the non-logarithmic case.
Of course, varying relative line profile size and line blending will make the logarithmic line profiles vary
in absolute size.
Nevertheless there is some equalization in line profile appearance in using logarithmic flux.
Why do we prefer the more equal appearance of plotting logarithmic flux when
large line features, of course, represent larger effects?
The reason is that small line features present
fine tests for spectrum modeling and clues about quantities (e.g., composition) that may not be
apparent in the large line features.
Thus, giving more equal weighting to large and small line features makes sense.

     Using local normalization has a similar effect to using logarithmic flux:  there is
an equalization in plotted size of line profiles of the same size relative to the continuum in
the ordinary spectra.
Nevertheless, we plot the logarithmic flux of the locally normalized spectra in plotting.
This is for consistency with our usual practice with ordinary spectra.
As mentioned above, in making locally-normalized spectra, we actually use logarithmic flux
and logarithmic wavelength in creating fitted lines for the large-scale smoothing rather than
just using flux and wavelength. 
This procedure in creating the locally-normalized spectra tends to make the locally-normalized continuum more
like 1 everywhere to the eye on a log-log plot.
% The logarithmic flux and logarithmic wavelength of the spectra are actually used (as mentioned above) in making the
% smoothed spectrum when we fit 
% a line to logarithmic flux as a function of logarithmic wavelength over the box-car interval
% (i.e., we fit flux with a power-law function over the box-car interval).
Locally-normalized spectra calculated using a line fitted to flux as a function of wavelength
% for the large-scale smoothing  % not needed
(which we will call the unlogged locally-normalized spectra) look
similar to those calculated with our logarithmic fitting procedure, 
but the difference is clearly significant to the eye. 

     Before applying the large-scale smoothing in our local normalization procedure, we also apply to all our spectra
a small-scale box-car smoothing (using trapezoid rule integration) with a relative smoothing length of
$\delta\lambda/\lambda=1/300$ which corresponds to a Doppler shift velocity of about $1000\,{\rm km\,s^{-1}}$.
This smoothing reduces the noise in some of the spectra particularly at the spectrum ends.
For some particularly noisy spectra, we sometimes use a larger small-scale smoothing length. 
The small-scale smoothing should not degrade the spectrum features very much.
Supernova line features generally vary relatively slowly over wavelength scales % shifts? scales is OK.
corresponding to Doppler shift velocities of order $1000\,{\rm km\,s^{-1}}$.
Interstellar and telluric lines can vary rapidly on shorter wavelength scales, but these are not intrinsic to
supernova spectra and could be eliminated from the spectra if necessary.

    A stand-alone fortran-95 code locnorm.f for making locally normalizing spectra 
is available for downloading from SUSPEND under the heading {\it Useful Programs}.
All the supernova spectra currently in SUSPEND are shown plotted in
original and locally-normalized format under the heading {\it Supernovae by Epoch} in html files
in supernova directories.

\section{EXAMPLES OF LOCALLY-NORMALIZED SUPERNOVA SPECTRA}

      Figures~1, 2, 3, and~4 (which we discuss individually below in separate subsections)
show locally-normalized supernova spectra plotted on log-log plots together
with the original spectra in the wavelength representation (i.e., $f_{\lambda}$ or flux-per-unit-wavelength
representation):
the locally-normalized spectra are obviously the ones with continuum levels of about 1 
and the original spectra have been vertically
shifted to be well displayed.
The spectra are all for Type~IIb supernova SN~1993J which occurred in 
\object{M81} (NGC~3031).
They have been deredshifted using host galaxy heliocentric velocity $-39\pm2\,{\rm km\,s^{-1}}$ 
from Leda \citep{paturel2003}.
(Since the host galaxy heliocentric velocity is negative, the spectra are, of course, actually redshifted to be
put in the rest frame.) 
The original spectra have not been dereddened.
The $E(B-V)$ reddening value for SN~1993J is quite uncertain.
\citet{richmond1994} after a lengthy consideration of many methods of determining $E(B-V)$ for 
SN~1993J suggest that
the SN~1993J $E(B-V)$ value is in the range $\sim 0.08$--$0.32\,$mag which is a range from low to moderate $E(B-V)$.
The possible intrinsic spectra obtained by correcting an observed spectrum using $E(B-V)$ values from this
range and a standard reddening law \citep[e.g.,][]{cardelli1989,o'donnell1994} would show a considerable
range in continuum behavior.
Thus, the intrinsic continuum behavior of SN~1993J for any phase is quite uncertain.
Avoiding the uncertainty in dereddening and calibrating
observed spectra is one of the main reasons for using local normalization
as discussed in \S~\ref{introduction}. 
% If one adopted this range for $E(B-V)$, the intrinsic SN~1993J color $B-V$ would be assigned 
% an uncertainty of about $\pm 0.12\,$mag.
% With this much uncertainty in 
% i.e., a range of color of .24. 
% See \citep[e.g.,][p.~498]{shore2003} for B-V=(B-V)_0+E(B-V).

    Because M81 is at a distance of only $3.63\pm0.34\,$Mpc (as determined from Cepheids \citep{freedman1994}), SN~1993J became
a very bright and well-observed supernova.
It was discovered on 1993 March 28.86 UT (JD~2449075.36) \citep{ripero1993}  % Just what Ripero says for visual.
which was probably only about a day 
after explosion (or core collapse) which may have been 1993 March $27.9\pm0.2$ (JD $2449074.4\pm0.2$) 
\citep{richmond1994}. % clocchiatti1995}.  
%%% richmond p. 1029 and clocchiatti p. 175, but the clocchiatti doesn't agree.
An initial very high peak in UVOIR bolometric brightness was partially observed \citep[e.g.,][]{richmond1994}:  
this peak, which probably happens for all core-collapse
supernovae soon after shock break-out, is usually unobserved, and so is not 
used to define conventional UVOIR bolometric maximum light even
though it may often/always be higher than conventional UVOIR bolometric maximum light as it is for SN~1993J.
SN~1993J's (conventional) maximum lights 
occurred on 
1993 April 16 (JD 2449093.7) in $B$, 
1993 April 17 (JD 2449095.0) in $V$, 
and 
1993 April $17\pm1$ (JD $2449094.7\pm1.0$) 
in UVOIR bolometric luminosity \citep{richmond1994}.   % See richmond 1033 & 1035.
The UVOIR bolometric luminosity rise time to maximum light was about 20~days \citep{richmond1994}. 
The maximum $B$ and $V$ apparent magnitudes were, respectively, $11.35\pm0.05$ and $10.80\pm0.05$
\citep{richmond1994}.  % p. 1033.

    SN~1993J was originally classified as a Type~II supernova because it had conspicuous hydrogen Balmer lines.
However, these lines did not become as strong as in typical Type~II supernovae and by about 40~days after explosion
SN~1993J evolved to resemble Type~Ib supernovae \citep[e.g.,][]{filippenko1993}.
Type~Ib supernovae have metal lines and conspicuous helium lines and, it seems in general, 
hydrogen lines that are somewhat weak and hard to identify especially after maximum light \citep{branch2002}.
Supernovae that undergo a transformation from Type~II to Type~Ib have come to be called Type~IIb's using 
a term invented by
\citet{woosley1987} in a theoretical prediction that small-hydrogen-envelope supernovae 
could exist.  % See Woosley p. 669. 
Type~IIb's are considered to be hydrogen-deficient compared to ordinary Type~II supernovae, but 
probably have more hydrogen than Type~Ib/c's.
The observational prototype Type~IIb is \object{SN 1987K} \citep{filippenko1988}.
Two other Type~IIb supernovae with a significant number of published spectra are
\object{SN 1996cb} \citep{qiu1999,matheson2001} and \object{SN 1998fa} \citep{matheson2001}. 
As of 2007 March 8, the supernova list of \citet{cbat2007} contains 25 Type~IIb or possible Type~IIb supernovae
(including SN~1987K which is actually listed as a Type~II).

    The ejecta of SN~1993J had a shock interaction with a thick circumstellar wind shed by the supernova
progenitor as evidenced by radio and X-ray emission \citep[e.g.,][and references therein]{fransson1996}.
In this respect, SN~1993J was like the Type~II supernovae \object{SN 1979C} and \object{SN 1980K} which also
showed strong radio emission \citep[e.g.,][]{weiler1986}.
The interaction also had an effect on the UV spectra of SN~1993J similar to that on SN~1979C and SN~1980K.  
We will discuss this UV effect briefly in \S~3.2.

     We have made some line identifications in the figures by labels giving the line's parent ion or
specific line name in the case of the Balmer lines. 
The labels are only given for the lines in the locally normalized spectra since the corresponding lines in the
original spectra are identified by comparison to the locally normalized spectra. 
For clear P~Cygni lines, we put the identification labels with the blueshifted absorption features of the P~Cygni lines
since this is usually the most distinct part of the line profiles. 
For emission lines, we put the identification labels with the emission features of the emission lines.
P~Cygni lines dominate optical/IR spectra in the photospheric phase and emission lines
in the nebular phase. 
Figures~1 and~2 show photospheric phase spectra and Figures~3 and~4 show nebular phase spectra.

     We should remark that in the photospheric phase, a supernova is optically thick in the optical/IR.
As the supernova expands, the opacities must fall with decreasing density and the supernova gradually
enters the nebular phase where the supernova is optically thin in the optical/IR.
Supernovae tend to remain
optically thick in the UV for much longer than in the optical/IR because 
there are many strong metal lines in the UV particularly
from the iron-peak elements.
There is no sharp dividing time between the two phases.
One often just starts using the term nebular phase when the emission features start to dominate
the optical/IR line pattern.
There is also no sharp distinction between P~Cygni lines (possessing blueshifted absorptions and 
emission features usually centered on the rest-frame line-center wavelengths)
and emission lines (which are just emission features usually centered on
the rest-frame line-center wavelengths). 
(The centering of emissions on the rest-frame line-wavelengths is often approximate and
sometimes the emissions are significantly shifted.
For example, strongly blueshifted emission lines can occur in the UV.  
For examples of such emission lines, see \S\S~3.2--3.4.)
In practice, when the absorption feature of a P~Cygni line has become hard to identify, we
can say it has become or nearly become an emission line.

    We should also remark that supernova matter after the explosion phase is in uniform motion and the whole
ejecta structure just scales up linearly with time $t$ since the explosion which is effectively a point event.
This motion is called homologous expansion.
In homologous expansion, the radial velocity of any matter element of ejecta is a good comoving-frame radial coordinate and we use it
as such as is customary in supernova studies.
The densities of the matter elements in the ejecta scale as $t^{-3}$.

\subsection{Figure~1: the 1993 March 31 Spectrum}

     Because of its closeness and brightness, SN~1993J was spectroscopically very well observed with the earliest
spectrum taken on 1993 March 29.88 UT (JD~2449076.38) with the Isaac Newton Telescope by E.~Perez and D.~Jones 
\citep{clocchiatti1995}.  % p. 169.  Julian date confirmed with calculator.
The spectrum in Figure~1 is also an early one from 1993 March 31 (JD 2449078.35) \citep{barbon1995} 
which is about 
4~days after explosion and is about 16~days before UVOIR bolometric maximum light.
The original spectrum rises very steeply to the blue showing that the supernova photosphere was very
hot at this phase.
From Figure~11 of \citet{richmond1994}, we estimate that a blackbody spectrum fit to the spectrum 
(if it was well calibrated) for this phase
would give a blackbody temperature in the range $\sim 12000$--$23000\,$K
where the range is caused by the uncertainty in the reddening.  % by far the dominant cause of range.
%% \citet{richmond1994} temperatures were based on fits to colors.
The uncertainty in spectrum temperature highlights the difficulty in relying on 
continuum shape for determining the physical properties of supernovae from  heterogeneous types
with uncertain reddening and supports our argument for giving more weight to the line pattern
than to the continuum shape (see \S~1).  % see between 1st and 2nd reason. 

     We have identified the lines in Figure~1 mostly following \citet{barbon1995} and \citet{baron1995}.
For this spectrum and the others below, we also rely on our general understanding of spectrum formation in
making identifications.
The H$\delta$ identification is uncertain because the absorption feature
identified as possibly due to H$\delta$ is rather weak and indefinite.
The He~I line is He~I $\lambda5875.7$ \citep[e.g.,][p.~14]{wiese1966}
and the Ca~II line actually arises from the Ca~II IR triplet
(i.e., Ca~II $\lambda8579.1$, where we have cited the $gf$-weighted mean line wavelength:  
the component lines are at $8498.02\,$\AA, $8542.09\,$\AA, and $8662.14\,$\AA)
\citep[e.g.,][p.~251]{wiese1969}.
(Hereafter we usually cite only the $gf$-weighted mean line wavelength for multiplets.)
The line velocities (i.e., the velocities corresponding to the Doppler blueshifts of the line absorption
minima from the rest-frame line-center wavelengths) are about 
$11000\,{\rm km\,s^{-1}}$ (H$\delta$ if this is the correct identification),
$5900\,{\rm km\,s^{-1}}$ (H$\gamma$),
$13400\,{\rm km\,s^{-1}}$ (H$\beta$),
$12800\,{\rm km\,s^{-1}}$ (He~I),
$12700\,{\rm km\,s^{-1}}$ (H$\alpha$),
and
$26800\,{\rm km\,s^{-1}}$ (Ca~II IR triplet).
We use the $gf$-weighted mean wavelengths for calculating the line velocities of multiplets.
(Here and throughout our discussion, wavelength measurements are made from the locally-normalized spectra and 
we do a little smoothing by eye to determine the minima when needed.) 
The low velocity of H$\gamma$ may be because of line blending with H$\delta$ since the lines
are separated in Doppler shift velocity by only $17000\,{\rm km\,s^{-1}}$.
The much higher velocity for the Ca~II IR triplet line than for the other lines makes the identification
a bit uncertain, but there seems nothing else it could be.
%% But where are the Ca II H&K lines???   Lost in calibration at the spectrum end?
In fact, high-velocity line features, for lines including the Ca~II lines, have been found 
for at least one other HDCC supernova
(Type~Ibc SN~2005bf \citep{folatelli2006,parrent2007};  see \S~\ref{introduction} for HDCC supernovae) 
and the work of \citet{parrent2007} imples that high-velocity line features might be
common in Type~Ic supernova spectra.   % See parrent last paragraph of text and S 4.  Not explicit.
(By high-velocity features, we mean line absorptions forming well above the photosphere at velocities
of perhaps $\gtrsim 15000\,{\rm km\,s^{-1}}$.)
Given the above remarks, it is possible, speaking speculatively, 
that HDCC supernovae of all kinds (such as Type~IIb SN~1993J) might have
some high-velocity line features in some cases.
% {branch2002} uses some detached H and He, but only one case with really HV

    A remarkable feature of the line pattern is that the H$\beta$ line is much stronger than the H$\alpha$ line:
this cannot happen in local-thermodynamic equilibrium (LTE).
The feature may require a non-LTE (NLTE) explanation.
On the other hand, the feature may be a result of an observational error of some kind since another 
SN~1993J spectrum
from the same day reported by \citet{baron1995} does not clearly show the H$\beta$ line stronger 
than the H$\alpha$ line.
In any case, \citet{baron1995} using the NLTE code PHOENIX were able to fit the continuum well,
but not the lines for the March~30--31 phase of SN~1993J.
Their difficulty may lie in the inaccurate physical structure of their model and/or an inaccurate
reddening correction.

\subsection{Figure~2:  the 1993 April 15 Spectrum}

      Figure~2 shows the SN~1993J spectrum from 1993 April~15 \citep{jeffery1994}:  this was
2~days before UVOIR bolometric maximum light and about 18~days after explosion.
The spectrum is a combination of an {\it HST} UV-blue-optical spectrum
and Lick Observatory ground-based spectrum.
The {\it HST} spectrum was obtained as part of the {\it HST} Supernova INtensive Study (SINS) General
Observer program \citep{kirshner1988}.
The two spectra agreed well in shape and scale in their overlap region $3120$--$3276\,$\AA, and so we simply
%%% the html file says the scales agreed.
joined them at $3240\,$\AA\ cutting off the overlapping ends.
The line identifications mostly follow from \citet{jeffery1994} and \citet{baron1995}.
The line velocities are in the range $\sim 7000$--$11000\,{\rm km\,s^{-1}}$ \citep{jeffery1994}.
We emphasize though that the lines are often blended and the identifications
in many cases only indicate the strongest contributor to the line feature. 
The 4 strongest hydrogen Balmer lines, the He~I $\lambda5875.7$ line and the Ca~II IR triplet line 
are again present.
We note that H$\alpha$ is now the strongest line in the optical part of the spectrum
(but note that the Mg~II line in the UV is comparably strong: see below)
and that it is in net emission which is typical of H$\alpha$ lines in Type~II supernovae.

    The absorption centered at about $3835\,$\AA\ is caused by the Ca~II~H\&K lines:  i.e.,
the multiplet Ca~II $\lambda3945.2$ \citep[e.g.,][p.~252]{wiese1969}.
%(citing the $gf$-weighted mean wavelength \citep[e.g.,][p.~252]{wiese1969}).  % redundant with section 3.1
This multiplet is very strong because it is a resonance multiplet (i.e., it arises from the ground level).
The line velocity for the Ca~II~H\&K lines is $8360\,{\rm km\,s^{-1}}$.
The Fe~II lines mainly arise from the multiplets Fe~II $\lambda4555$, Fe~II $\lambda4561$, 
and Fe~II $\lambda5060$ \citep[e.g.,][Table~4]{kirshner1993}.
% (citing the $gf$-weighted mean wavelengths \citep[e.g.,][Table~4]{kirshner1993}).
The absorption centered at about $7570\,$\AA\    % (7538 + 7601)/2 = 7570
may be the absorption of a P~Cygni line caused by
the multiplet O~I~$\lambda7773.4$ \citep[e.g.,][p.~79]{wiese1966}.
This line is common in Type~Ib/c supernovae \citep[e.g.,][]{branch2002} and Type~Ia supernovae
\citep[e.g.,][]{kirshner1993}, but unfortunately its absorption usually coincides with a
strong telluric line and it is sometimes, as here, unclear if the telluric line has been adequately
corrected for.
We regard this identification as tentative. 
If the identification is correct, the O~I~$\lambda7773.4$ line velocity is $7830\,{\rm km\,s^{-1}}$.  % MgII7890 is too fast to be right.
By about 40~days past UVOIR bolometric maximum light, the O~I~$\lambda7773.4$ line seems clearly present
\citep{barbon1995}.

   The UV of SN~1993J is similar to that of SN~1979C and SN~1980K and is probably greatly modified
from what a bare supernova would have given by circumstellar shock interaction \citep{jeffery1994}. 
In a series of papers, \citet{fransson1982,fransson1984a,fransson1984b} and \citet{fransson1984c}
have given a detailed model of supernova UV behavior with circumstellar interaction:  this
work was stimulated by International Ultraviolet Explorer ({\it IUE}) data for  % stimulated was F's own word.
SN~1979C and SN~1980K.
In brief, some fraction of UVOIR light from the supernova photosphere is scattered by shock-heated electrons
in the shock region of ejecta and circumstellar matter and Comptonized (i.e., blueshifted up to the
UV or extreme UV).
Also there will be X-ray emission from the shock-heated gas.  % See fransson1984b, p. 268.
Some of the Comptonized light and X-ray emission just escapes the supernova environment, but some back-heats
the outer supernova ejecta and creates a complex, layered, photoionization region in the outer 
supernova atmosphere. 
The emission from back-heating adds to the direct emission from the supernova interior
that emerges at the photosphere.
In the optical, this emission from back-heating is probably negligible (except in some lines), but
in UV blueward of perhaps $\sim 2800\,$\AA\ it appears to become  %% increasingly 
important by comparison to spectra from supernovae with little or no circumstellar interaction
\citep{jeffery1994}.
As well as continuum emission from recombination and free-free emission, UV emission lines from
ionized atoms can make a considerable contribution:  the prominent UV emission lines are primarily
resonance lines that have been collisionally excited by the back-heated medium.  % fransson1984b, p. 274.
The layers of Comptonized-flux photonization form a geometrically narrow shell
of ejecta moving at velocities of order $10^{4}\,{\rm km\,s^{-1}}$.
The observer sees the emission from back-heating primarily from this shell. 
Below the shell is the optically thick photosphere.
Thus, the observer mostly sees the emission from back-heating from the near-side of the ejecta with 
line-of-sight velocities ranging from much less than $10^{4}\,{\rm km\,s^{-1}}$ away from
the observer up to line-of-sight velocities of order $10^{4}\,{\rm km\,s^{-1}}$ toward the observer.
For UV emission lines, 
the varying Doppler shift given by the moving shell will result in broad blueshifted emission features.
% The Comptonized flux dominates the line emission for of order 30~days after explosion and
% then X-ray emission from the circumstellar interaction begins to dominate.   % See fransson1984b, p. 272.
%% My reading of fransson1984b is too superficial.

     An effect that tends to enhance blueshifted line emission from a shell surrounding
an opaque photosphere is diffuse reflection toward the observer
of the line emission from the photosphere \citep{chugai1988}. 
This reflected line emission must mainly come from the near side of the photosphere,
and thus will tend to be mainly blueshifted from the rest-frame line-center wavelength and hence
the enhancement of the blueshifted line emission.
If the interior of the shell were transparent, the light that could otherwise  % "could" because there is absorption.
be reflected would simply stream away from the observer and all redshifted line emission
from the far side of the shell would come toward the observer. 
The reflection can be caused by scattering from electrons, but in the UV,
particularly in the nebular phase when the electron optical depth through the ejecta
has become small, the reflection is probably mainly by scattering from the many
iron-peak-element lines in the UV.
%  resonance lines or iron-peak-element lines arising from low-energy metastable levels. %% Not sure of this.
%  We know the UV line opacity is way above the electron scattering opacity:  harkness1990 in the ESO book.
To understand the reflection effect clearly, imagine an ideal case of a perfectly reflecting
photosphere just interior to a very geometrically thin line-emitting shell.
Virtually all the line emission from the near side of the shell moving inward would be reflected from the
near side of the photosphere and some would go to the observer.
Virtually no emission or reflection would come to the observer from, respectively, the far side of the shell or 
the far side of the photosphere.
Since the photosphere matter is expanding homologously, the reflected light toward
the observer will be mostly blueshifted.
Note that in moving inward toward the photosphere, the line emitted photons
will redshift from the rest-frame line-center wavelength in the comoving frame
and that the photosphere is moving a little more slowly than the shell.
(Propagating photons always redshift in the comoving frame in homologously expanding
atmospheres no matter what the degree of special relativistic effects \citep{jeffery1993}.)   % See jeffery1993-738. 
Consequently, the light reflected toward the observer
will tend on average to be less blueshifted in the observer frame than the line emission
coming directly from the line-emitting shell. 
Because the photosphere is just interior to the shell in our ideal case, this redshift effect will be small.
In a real case, the shell will not necessarily be very geometrically thin, and so some 
significant amount of
redshifted flux from the far side of the line-emitting shell could reach the observer.
Also some fraction of the observer-directed reflected light (mainly from near the 
photosphere limb where the
line-of-sight velocity of the photosphere is small) will end up a bit redshifted from the line-center wavelength
in the observer frame because of the redshift in the comoving frame that happens while 
the light moves inward toward the photosphere.
But the contribution of this redshifted reflected light is probably relatively small in most cases.
Note also, a real photosphere will not, of course,
be perfectly reflecting:  some of the line emission will be absorbed and
its energy dispersed to other parts of the spectrum.

     We have not attempted a full identification of features seen in the UV region of the
spectra we present. 
However, the strong emission-like feature peaking at $2717\,$\AA\ 
in the spectrum must be
the resonance multiplet Mg~II $\lambda2797.9$ (e.g., \citealt[p.~30]{wiese1969}; \citealt[p.~10]{moore1968})
blueshifted and in emission and produced as discussed by Fransson.
(The emission features most fully analyzed by Fransson were primarily those from blueward of $2000\,$\AA, but  
his analysis seems to apply to Mg~II $\lambda2797.9$ as well.)  % see Fransson 1984b p. 275, 278.
In SN~1979C UV {\it IUE} spectra \citep{ines2000}, a similar blueshifted Mg~II $\lambda2797.9$ emission 
appeared possibly about 7~days after optical maximum light 
(1979 April 22) \citep{panagia1980}, 
became definite by 33~days past optical maximum (1979 May 18), 
and was still clearly present in the last {\it IUE} observation about 111~days past maximum (1979 August 4).
(Optical maximum light for SN~1979C was about 1979 April 15 \citep{barbon1982}.) 
In SN~1980K UV {\it IUE} spectra \citep{ines2000}, a similar blueshifted Mg~II $\lambda2797.9$ emission 
seems to have arisen about 26~days after optical maximum light (1980 October 30) 
and be present on the last {\it IUE} observation from about 62~days past optical maximum
(1981 January 5).
% seem to have poorer quality than the SN~1979C spectra
In the SN~1980K spectra,
the emission is generally less clear than in SN~1979 spectra, but
it definitely present about 35~days past optical maximum (1980 December 9).  %  \citep{ines2000}. redundant now.
(Optical maximum light of SN~1980K was about 1980 November 4 \citep{buta1982}.) 
We note that an Mg~II $\lambda2797.9$ emission line is a common feature of quasar emission-line
clouds \citep[e.g.,][]{reichard2003} which may somewhat resemble the UV emission shell of circumstellar-interacting 
supernovae \citep{fransson1984b}.  % see fransson's p.~268.

   The Mg~II $\lambda2797.9$ emission peak wavelength $2717\,$\AA\ 
corresponds to a Doppler blueshift velocity of $8700\,{\rm km\,s^{-1}}$ 
which is within the range of P-Cygni line velocities in the optical $\sim 7000$--$11000\,{\rm km\,s^{-1}}$ 
\citep{jeffery1994}.
If we just define $2700\,$\AA\ as the characteristic blue edge of the Mg~II $\lambda2797.9$ emission (which is
good for later phases:  see \S\S~3.3 and~3.4 below), the
corresponding Doppler blueshift velocity is $10500\,{\rm km\,s^{-1}}$.
We take this velocity as the characteristic velocity of the Mg~II $\lambda2797.9$ emission shell.

   Without strong circumstellar interaction, the Mg~II $\lambda2797.9$ multiplet would probably contribute
along with Fe II lines to a blueshifted P-Cygni absorption feature.
For Type~Ia supernovae, the absorption can be seen in the {\it HST} spectra of \object{SN 1992A} in the photospheric
phase \citep{kirshner1993} and, perhaps, well into the nebular phase at 291~days after $B$ maximum light
although not so identified \citep{ruiz-lapuente1995}.
For Type~II supernovae, the absorption can be seen in the {\it IUE} spectra of SN~1987A \citep{pun1995}. 
The difference for the Mg~II $\lambda2797.9$ multiplet between having blueshifted emission and having blueshifted
absorption is probably explained by saying that with circumstellar interaction
the line-forming region is hotter than the inner ejecta and without circumstellar interaction it is colder. 

   No other identifications in the UV of SN~1993J are probably possible without a more detailed analysis.  
We leave that to future work {\it sine die}.

\subsection{Figure 3:  the 1993 September 15.5 Spectrum}

      Figure~3 shows the SN~1993J spectrum that we have constructed from
a 1993 September~17 {\it HST} UV-blue-optical spectrum and a 1993 September~14 ground-based spectrum
taken at the MMT Observatory by P.~Challis and C.~S.~J.~Pun.
We will call the combined spectrum the 1993 September 15.5 spectrum:  thus it comes
from about 151~days after UVOIR bolometric maximum light and 171~days after explosion. 
Both component spectra were obtained as part of the SINS program and neither
have been published before (\citealt[for the {\it HST} spectrum]{challis2006};
\citealt[for the MMT spectrum]{blondin2007}).  
% So Pete Challis (HST) and St\'ephane Blondin (optical) tell me.
The {\it HST} spectrum has not been given a definitive flux calibration \citep{challis1994}.
% and we are unsure about the ground-based spectrum.  % best say nothing.
The two spectra agreed well in shape in the region $4250$--$4781\,$\AA, and so we simply
% scaled them to the same overall level % redundant
and joined them at $4253.6\,$\AA\ cutting off the overlapping ends.
The ground-based spectrum had to be scaled by a factor of $0.35$.  %  See /sn1993j_Ia/trans2.f

     The supernova at the time of this spectrum is clearly in the nebular phase because of
the strong optical/near-IR emission lines. 
The line identifications for the 3 strongest emission lines in the optical/near-IR are based on
analysis of \citet{houck1996} for a 1993 August~15 SN~1993J spectrum:   the extrapolation of the
identifications to 1993 September 15.5 is
reliable because these lines are strong, persistent, and well identified in other HDCC supernovae.
The actual lines are the forbidden multiplet [O I] $\lambda\lambda6300,6364$ which is a resonance
multiplet \citep[e.g.,][p.~82]{wiese1966} and the forbidden multiplet [Ca II] $\lambda\lambda7291,7324$
\citep[e.g.,][p.~255]{wiese1969} and, again, the Ca~II IR triplet.

    The absorption with minimum at $3829\,$\AA\ is caused by Ca~II~H\&K lines 
(the resonance multiplet Ca~II $\lambda3945.2$).
The line velocity of the Ca~II~H\&K lines is $8800\,{\rm km\,s^{-1}}$.
There is no obvious strong emission feature associated with these lines. 
Evidently, the Ca~II~H\&K lines are optically thick and strongly trap photons in the 
spatial resonance regions of the lines.
The trapping process is as follows.
As photons are propagating through the supernova atmosphere,
they redshift into resonance with the lines in some location (i.e., a resonance region)
and are absorbed and then resonantly emitted, but cannot easily redshift 
out of resonance because of continually being re-absorbed and re-emitted in the lines
after traveling and redshifting very little.
So there are many re-absorptions and re-emissions.
Much of the photon energy ultimately escapes by being emitted by the Ca~II IR triplet which 
shares the same upper level with the Ca~II H\&K lines \citep[e.g.,][p.~12]{moore1968}:
this energy leak mostly suppresses the emission feature of the Ca~II~H\&K lines. 
The lower level of the Ca~II IR triplet is metastable and the upper level of the 
forbidden multiplet [Ca II] $\lambda\lambda7291,7324$ \citep[e.g.,][p.~12]{moore1968}.
Collisions are weak in the nebular phase due to low density, and hence the transitions
from the metastable lower level of the Ca~II IR triplet to the ground level
become increasingly radiative which results in the
strong [Ca II] $\lambda\lambda7291,7324$ emission feature.
The process of energy from absorption in the Ca~II H\&K lines being
emitted in other redder lines is an example of fluorescence \citep[e.g.,][p.~22]{mihalas1978}.

     In the UV, the blueshifted Mg~II $\lambda2797.9$ emission line now peaks at $2726\,$\AA\ which
corresponds to Doppler blueshift velocity of $7700\,{\rm km\,s^{-1}}$ and has now increased in height to
about $3.4$ times the continuum level.
Its overall shape, however, is much the same as earlier on 1993 April 15 (see Fig.~2 in \S~3.2
and Fig.~5 in \S~3.4).

     The peak centered at $4563\,$\AA\ may be at least in part the emission feature
of the semi-forbidden line Mg I] $\lambda4571$ (e.g., \citealt[p.~16]{moore1968};
\citealt[p.~26]{wiese1969};  \citealt{nist2006}).
The emission is relatively weak, but \citet{houck1996} predict Mg I] $\lambda4571$ as 
a weak emission line (seemingly blended with some other emission line)
for 1993 August~15 and believe the Mg I] $\lambda4571$ emission
should become stronger with time.  % see houck 815 and 816 and 822.
If the $4563\,$\AA\ emission is primarily Mg I] $\lambda4571$, then the line has been Doppler blueshifted
by velocity $\sim 500\,{\rm km\,s^{-1}}$.
The 1993~November 15.5 spectrum in Figure~4 (see \S~3.4) shows a stronger emission peaking at $4551\,$\AA\ and if this
is caused by Mg I] $\lambda4571$, then it has been Doppler blueshifted by velocity $\sim 1300\,{\rm km\,s^{-1}}$.
The 1993~November 15.5 spectrum emission peaking at $4551\,$\AA, however, has a secondary peak at
almost exactly $4571\,$\AA.
This suggests that the main peak is caused by some other line in emission and not Mg I] $\lambda4571$.
In order for the $4563\,$\AA\ and $4551\,$\AA\ emissions to be primarily Mg I] $\lambda4571$,
there must be some blueshifting asymmetry in the ejecta or some significant continuum optical depth.
(A significant continuum optical depth would tend to suppress the red side of an emission line.)
We note that the [O I] $\lambda6300$ emission peaks at $6302\,$\AA\ (corresponding to a
Doppler redshift velocity of $\sim 100\,{\rm km\,s^{-1}}$) in the  1993~September 15.5 spectrum and 
at $6300\,$\AA\ (corresponding to virtually no Doppler shift at all) in the  1993~November 15.5 spectrum.
The lack of significant shifts in the [O I] $\lambda6300$ emission peaks argues for symmetric
ejecta and optically thin ejecta in the optical.
There is the complication, however, that the blue side of the optical generally has more significant weak lines
than the red side of the optical and may persist being somewhat optically thick longer than the red side
due to a quasi-continuum opacity of the weak lines.
The upshot of all these considerations is that we draw no conclusion about 
the identities of the $4563\,$\AA\ and $4551\,$\AA\ emissions.
These emissions may be blueshifted Mg I] $\lambda4571$ or Mg I] $\lambda4571$ blended with some other
emission line, but more analysis is needed. 

    (The case for Mg I] $\lambda4571$ identification in the optical is different than
for the Mg~II $\lambda2797.9$ identification in the UV.
We expect the nebular-phase ejecta to be optically thin or relatively optically thin in the optical 
and optically thick in the UV at least for a long time into the nebular phase.
The UV optical thickness in the nebular phase 
is the cause of the blueshift of the Mg~II $\lambda2797.9$ emission as we argued in \S~3.2.)

     We note that the strong emission features in the optical/near-IR in both versions of the spectrum,
but particularly in the locally-normalized version, seem to be adjacent to absorptions.
There may be real strong absorptions at these locations from lines that are optically thick, but
a synthetic spectrum analysis is probably needed to show that.
The locally-normalized spectra for the nebular phase, however, may be a bit deceptive. 
The process of local normalization will tend to make emission features sit in troughs that are below 1, and
thus create the appearance of adjacent absorptions.  
In reality, the emission lines may just be added to a fairly smooth continuum that holds across the spectrum
(or at least the optical/IR spectrum) and there may be no significant absorptions (at least in the optical/IR
spectrum). 
So in regard to nebular spectra, local normalization as we have implemented it may not be ideal in that
it probably cannot scale a real physical nebular continuum to nearly 1 
at least in the vicinity of strong emission features.
On the other hand, there may be no well-defined physical continuum level in which case applying local normalization
does not obviously worsen the appearance of the spectra for analysis.
Our arguments for local normalization based on avoiding dealing with uncertain reddening 
and calibration error (\S~\ref{introduction})
are still valid for the nebular phase. 
Figure~3 and Figure~4 (\S~3.4) do show that local normalization even in the nebular phase
will, of course, tend to cause the average flux level to be near 1.

     For the photospheric phase, local normalization is probably much better at
setting the phyisical continuum to near 1 than for the nebular phase. 
First, there is a significant real physical continuum in the spectrum (although it is probably
almost never a pure blackbody continuum) which is not necessarily the case in the nebular phase.
Second, the smoothing of the P~Cygni absorptions and emissions to create the large-scale smoothed spectrum
is an averaging that must put the large-scale smoothed spectrum close to the physical continuum level.
The locally-normalized spectrum will then have a continuum close to 1.

\subsection{Figure 4:  the 1993 November 15.5 Spectrum}

      Figure~4 shows the SN~1993J spectrum that we have constructed from
a 1993 November~14 {\it HST} UV-blue-optical spectrum and a 1993 November~17 Lick Observatory
ground-based spectrum both of which were obtained as part of the SINS program.
The {\it HST} spectrum has not been published before \citep{challis2006};     % Pete Challis so informs me.
the Lick spectrum was taken by A.V.~Filippenko and T.~Matheson \citep{filippenko1994}.
We will call the combined spectrum the 1993 November 15.5 spectrum:  thus it comes
from about 212~days after UVOIR bolometric maximum light and 232~days after explosion.
The {\it HST} spectrum has not been given a definitive flux calibration
\citep{challis1994} and the ground-based spectrum is not absolutely calibrated, although its 
relative calibration is excellent \citep{filippenko1994}. 
The two spectra agree well in shape in the overlap region $3120$--$4781\,$\AA, and so we simply
% scaled them to the same overall level and  % redundant
joined them at $4200\,$\AA\ cutting off the overlapping ends.
The ground-based spectrum,
after converting it from the frequency representation to the wavelength representation,
had to be scaled by a factor of $5.55\times10^{-9}$ to account for overall calibration
differences and for unit transformation.
% Alex says in mJy.  See /sn1993j_Ia/trans2.f 
The {\it HST} spectrum blueward of $\sim 2425\,$\AA\ seems to decline in a manner too steep to be physically
real and we do not trust it (nor the locally-normalized spectrum we derive from it) there.

     Qualitatively, the 1993 November 15.5 spectrum is much like the 1993 September 15.5 spectrum.
(We are relying on the locally-normalized spectra for comparisons.)
The Mg~II $\lambda2797.9$, [O I] $\lambda\lambda6300,6364$, and [Ca II] $\lambda\lambda7291,7324$
emissions have increased in height relative to the continuum and the Ca~II IR triplet emission has decreased.
The peak of the Mg~II $\lambda2797.9$ emission is at $2726\,$\AA\ just as in the 1993 September 15.5 spectrum.
The emission at peaking at $4563\,$\AA\ (whatever it may be) in the 1993 September 15.5 spectrum has increased and
shifted so that it now peaks at $4551\,$\AA. 
(See the discussion of this emission feature in \S~3.3.)
The Ca~II H\&K absorption is not as deep as in the 1993 September 15.5 spectrum.
 
     In Figure~5, we show the time development of Mg~II $\lambda2797.9$ emission in locally-normalized spectra.
Although the emission line scales up with time, its appearance is remarkably stable.
Our definition of $2700\,$\AA\ as the characteristic blue edge of the Mg~II $\lambda2797.9$ emission (with 
corresponding Doppler blueshift velocity $10500\,{\rm km\,s^{-1}}$;
see \S~3.2) is valid for all the phases we display.
The behavior of the spectra in the region $\sim 2780$--$2820\,$\AA\ is not very certain since we have
just omitted wavelength points that correspond to the interstellar Mg~II $\lambda2797.9$ lines which
arise from gas clouds along the line of sight to the supernova.
Such clouds, identified from interstellar lines of various ions, have
heliocentric velocities in the range from about $-140$ to $181\,{\rm km\,s^{-1}}$
\citep{wamsteker1993,wheeler1993}.   % Cited in jeffery1994.

     Later {\it HST} UV spectra of SN~1993J from day~$+649$ to about day~$+2561$ relative
UVOIR bolometric maximum light
show the Mg~II $\lambda2797.9$ emission as box-like in appearance and more symmetrical
about the rest-frame line-center wavelength than in the spectra we show \citep{fransson2005}.
The Mg~II $\lambda2797.9$ emission in the later spectra have 
a Doppler velocity full-width of $\sim 17000\,{\rm km\,s^{-1}}$.
It seems that the circumstellar interaction back-heating persists through the later phases,
but the optical depth through the ejecta has greatly diminished, and 
thus occultation of the far side of the Mg~II $\lambda2797.9$ shell has greatly decreased. 
% fransson2005 posits.  See his 994, 995, 1002.
The flattish top of the late  Mg~II $\lambda2797.9$ emission line is 
characteristic of emission lines from expanding, optically
thin shells as first shown by \citet{beals1931}:  other references for the flat-top emission from
expanding shells are \citet[p.~477]{mihalas1978} and \citet[p.~190]{jeffery1990}.

\subsection{The Continuum Independence of Local Normalization}

    The main goal of local normalization is to eliminate variations in the continuum in comparing spectra. 
Implicit in this goal is the requirement that locally-normalized spectra be virtually independent of
the continuum shape.
For example, if this were not the case, intrinsic spectra that were identical, but with
observed continuums distorted in various ways, could be transformed
to different locally-normalized spectra and one would reach the false conclusion that the intrinsic spectra were 
not identical.
To test local normalization for continuum independence, 
we have weighted each of the four original spectra in this section by $\lambda^{2}$
(effectively converting the spectra to the $f_{\nu}$ representation)
and $\lambda^{4}$ and then locally normalized these weighted spectra and compared them to each other and the
unweighted locally-normalized spectra. 

    For each original spectrum, a plot (using an expanded scale compared to that spectrum's
plot in Fig.~1--4)
of the corresponding locally-normalized spectra calculated with and without weights 
shows exact overlap to the eye of the locally-normalized spectra. 
On the other hand, the unlogged locally-normalized spectrum (see \S~2.2) 
%% See 3rd to last paragraph of 2.2 for unlogged locally-normalized spectra.
corresponding to an original spectrum
was distinct from the other locally-normalized spectra 
for that original spectrum 
when plotted:  
the unlogged locally-normalized spectrum was generally a bit lower than the other locally-normalized spectra.

    As mentioned in the Introduction (\S~1), DIFF1 is essentially a mean relative difference between
the line patterns of two spectra, and thus the reader can roughly understand DIFF1 values before seeing
the precise formula (which is given below in \S~4.1). 
The DIFF1 values for spectrum pairs consisting of the locally-normalized weighted spectra and 
the locally-normalized unweighted spectra (all derived from the same original spectrum)
were of order $0.001$--$0.0015$ for weight $\lambda^{2}$ and of order $0.002$--$0.003$ for weight $\lambda^{4}$.
These are very small mean relative differences.
On the other hand, DIFF1 tests of the unlogged locally-normalized spectra with the corresponding
other locally-normalized spectra
gave DIFF1 values of order $0.4$--$0.7$ which are comparable to DIFF1 values between 
locally-normalized spectra of the same
supernova from phases differing by several days.   % See sn1993j_04_15 for a good example.

    We conclude that local normalization will transform continuum-distorted versions of an 
original spectrum to virtually the same locally-normalized spectrum.
Thus, local normalization is sufficiently continuum-independent
to lead to valid results in comparing locally normalized spectra.
On the other hand, different local normalization procedures can produce noticeably distinct 
locally-normalized spectra for a given original spectrum.
Thus, one should choose a single local normalization procedure and stick with it for valid
comparisons of locally-normalized spectra.

\section{DIFF1 AND DIFF2}

     In this section, we introduce the formulae for DIFF1 (\S~4.1) and DIFF2 (\S~4.2), show 
an example of two locally-normalized spectra fitted by
minimizing the DIFF2 function value (\S~4.3), and compare DIFF1 and DIFF2 (\S~4.4).

\subsection{DIFF1}

     Say $f_{i}$ is the locally-normalized flux at wavelength $\lambda_{i}$. 
Let 
\begin{equation}
\delta_{i}=f_{i}-1
\label{eq-flux-difference}
\end{equation}
be the flux difference at wavelength $\lambda_{i}$.
The flux difference is the relative difference of the line pattern from the continuum for the
original spectrum or the absolute
difference of the line pattern for the locally-normalized spectrum.
We define DIFF1 between a spectrum~1 and a spectrum~2 by the formula 
\begin{equation}
{\rm DIFF1}={1\over I}\sum_{i}{ | \delta_{1,i}-\delta_{2,i} |
                               \over \max\left(|\delta_{1,i}|,|\delta_{2,i}|,\xi\right)} \,\, ,
\label{eq-DIFF1}
\end{equation}
where the subscripts 1 and 2 indicate the spectrum and $I$ is the total number of wavelength points
in the summation.
(The wavelength points are equally spaced in logarithmic wavelength:  see below.)
The $\xi$ is a small number that we set to $10^{-15}$.
This is typically the smallest number that added to 1 that creates a number significantly
different from 1 with fortran~95 double precision arithmetic.
The $\xi$ prevents arithmetic exceptions, but is only very rarely invoked. 

    From equation~(12), it is clear why we describe DIFF1 as the mean relative difference in 
line pattern between two spectra (see \S\S~1 and~3.5).
If ${\rm DIFF1}<<1$, the spectra will be much alike.
If ${\rm DIFF1}\gtrsim 1$, the spectra will be substantially different.
For example, if the flux differences (i.e., the $\delta_{i}$'s)
for one spectrum in a DIFF1 test were almost all only a small
fraction of those for the other spectrum (i.e., one spectrum had much weaker lines than the other),
the DIFF1 value would be close to 1.
In fact, the DIFF1 function has an upper limit:
\begin{equation}
 | \delta_{1,i}-\delta_{2,i} |\leq |\delta_{1,i}|+|\delta_{2,i}|
   \leq 2\max\left(|\delta_{1,i}|,|\delta_{2,i}|\right) 
   \leq 2\max\left(|\delta_{1,i}|,|\delta_{2,i}|,\xi\right) \,\, ,
\end{equation}
and thus
\begin{equation}
{\rm DIFF1}\leq 2 \,\, .
\end{equation}
We would not usually expect a DIFF1 value approaching 2 since that would mean that two input 
locally-normalized spectra
had line patterns that were nearly mirror images of each other about the continuum level:
i.e., the line patterns would be strongly anticorrelated.
In fact, any DIFF1 values well over 1 would show some anticorrelation between spectrum line patterns.
There is no strong physical reason to expect much anticorrelation of supernova spectrum line patterns,
but some accidental anticorrelation should happen, and so some DIFF1 values over well over 1 may occur when
applying DIFF1 to heterogeneous samples of spectra.

    We have subjected the set of all possible spectrum pairs drawn from the sample of HDCC supernovae described in \S~5
to the DIFF1 test.
(There are 17 supernovae, 168 spectra, and 27588 valid spectrum pair tests:  some possible pair tests are excluded
as invalid by the validity criterion discussed below.)  
The resulting distribution of DIFF1 values has
mean, standard deviation, minimum value, and maximum value of, respectively, 0.865, 0.137, 0.244, and 1.320
(where we have reported more digits than are significant to allow for numerical consistency checks). 
That the mean 0.865 is so close to 1 (actually within $1\sigma$ of 1)
shows that spectrum pairs on average are not highly similar.
This is understandable since the spectra come from a heterogeneous sample of supernovae and from
many phases:  the spectrum of a supernova can evolve significantly with phase.
The standard deviation value shows that spectrum pairs significantly (i.e., $1\sigma$)
more alike than average
will have DIFF1 values of $\lesssim 0.73$. 
The minimum 0.244 suggests that even spectra from the same supernova at close to the same
phase vary from each other significantly. 
Note we have excluded redundant spectra from the same phase period (which we usually set to being 1 day)
for a supernova (as described 
in \S~5.1) since those should be nearly identical to the included spectrum aside from observational error.
% See 5.1, para 2.
As we expected, the maximum DIFF1 value is much less than 2.
 
    We note that if we used $f_{i}$'s in the formula for DIFF1 (i.e., eq.~(12)) instead of $\delta_{i}$'s
(which would only change denominators, in fact),
we would have lessened the sensitivity to line patterns.
For example, Type~IIb and Type~Ic~hypernovae at early phases (i.e., well before maximum light)
both have relatively weak lines in the optical.
In the case of Type~IIb supernovae, this is because they have hydrogen-dominated atmospheres that
when hot (i.e., over of order $10000\,$K) show weak lines. 
(Why the lines should be as weak as they are in early Type~IIb spectra has not yet
been fully elucidated theoretically by NLTE analysis \citep{baron1995}.)  % see baron1995, p. 174, right column
Type~Ic hypernovae show weak lines at early times because they are hot (which tends to cause
weak optical lines also in cases with little or no hydrogen) 
and because their velocity
structure is very fast giving them extremely broad P-Cygni lines with line velocities of
up to $30000\,{\rm km\,s^{-1}}$ and perhaps higher \citep[e.g.,][]{mazzali2000,patat2001}.
The stretching out of P-Cygni lines by very high velocities tends to make them shallower and more overlapping.
Both effects tend to give line features a small vertical scale relative to the continuum.
If we used $f_{i}$'s in the formula for DIFF1 instead of $\delta_{i}$'s, the DIFF1 values
between early phase spectra of Type~IIb's and Type~Ic~hypernovae would be small even though
the spectra and the supernovae are intrinsically quite different because the 
formula numerators would be relatively small and the formula denominators relatively large. 
However, the given formula for DIFF1 distinguishes the two types of spectra 
because the formula denominators
become small when the lines features are small.
%% Probably types of spectra is better than types of supernovae in this case.
%% Maybe it's a toss-up. 

    For our calculations, we interpolated the locally-normalized spectra onto a grid of
equally spaced points in (natural) logarithmic wavelength.
For coding convenience, we chose $\ln(1\,${\rm \AA}$)=0$    % A stupid form, but it prevents a latex warning.
to be the zero point of the logarithmic wavelength scale. 
For the grid increment, we chose 
\begin{equation}
\Delta\ln(\lambda)={1\over 3000} \,\,  .
\label{eq-log-lambda-grid-increment}
\end{equation}
If the grid wavelengths labeled by $i$ are $\lambda_{i}$ and we define
$\Delta\lambda_{i}=\lambda_{i}-\lambda_{i-1}$, we find that
\begin{equation}
\Delta\ln(\lambda)=\ln(\lambda_{i})-\ln(\lambda_{i-1})=\ln\left({\lambda_{i}\over\lambda_{i-1}}\right)
                  =\ln\left(1+{\Delta\lambda_{i}\over\lambda_{i-1}}\right)
                  \approx {\Delta\lambda_{i}\over\lambda_{i-1}} 
\end{equation}
to first order in small ${\Delta\lambda_{i}/\lambda_{i-1}}$.
We now see that the logarithmic wavelength increment corresponds 
to a relative change in wavelength for each increment of approximately $1/3000$.
Thus, near $3000\,$\AA, the wavelength increment is about $1\,$\AA;  near $6000\,$\AA, about $2\,$\AA.
The Doppler shift velocity corresponding to a relative wavelength increment of $1/3000$ is about $100\,{\rm km\,s^{-1}}$.
As mentioned in \S~2.2,  % See 2nd to last paragraph of the section.
supernova line features generally change relatively slowly over wavelength shifts corresponding
to velocity changes of order $1000\,{\rm km\,s^{-1}}$, and thus our grid should be sufficiently fine for almost
all supernova spectra.
As also mentioned in \S~2.2, interstellar and telluric lines can vary on 
shorter wavelength scales, but these are not intrinsic to
supernova spectra and could be corrected for if necessary. 

      We only apply the DIFF1 test to what we call the good overlap region 
in logarithmic wavelength of two locally-normalized spectra.
In this logarithmic wavelength region, both the spectra seem physically good.
By physically good, we mean not excessively noisy and not showing unphysical behavior in a narrow 
wavelength band because of some calibration error.
We only allow one good region for a spectrum:  i.e., we never multiple good regions for
a spectrum which are separated in wavelength.
This is not much of a limitation since the bad regions are usually the
spectrum ends which can often be very noisy and have narrow wavelength band calibration errors.
We only consider DIFF1 tests valid if the good overlap region in logarithmic wavelength
% (good in the sense just specified above)  % redundant
is $\geq 0.2$ which corresponds to a good overlap region in relative wavelength
of $\sim 22\,$\% relative to the lower wavelength of the region.
Since almost all the spectra we consider are good or nearly good over the range $4000$--$7000\,$\AA\
(logarithmic wavelength range $\sim 0.5596$ or relative wavelength range of $75\,$\% relative to $4000\,$\AA), 
almost all DIFF1 evaluations are considered valid.

\subsection{DIFF2}

      The DIFF2 formula is
\begin{equation}
{\rm DIFF2}={1\over I}\sum_{i}{ | \delta_{1,i}-\delta_{2,i+k} |
                               \over \max\left(|\delta_{1,i}|,|\delta_{2,i+k}|,\xi\right)} \,\, . 
\label{eq-DIFF2}
\end{equation}
The only difference from equation~(\ref{eq-DIFF1}) for DIFF1 (\S~4.1) is the term $k$ in the 
wavelength subscript of one of the flux differences (defined by eq.~(\ref{eq-flux-difference}) in \S~4.1).
This $k$ is chosen to minimize the DIFF2 function value:   this minimized value is the DIFF2 value itself.
We allow up and down (natural) logarithmic wavelength shifts of only up to $0.05$.
Thus, we allow up and down relative wavelength shifts of only about $5\,$\%.
Given that the logarithmic wavelength grid increment is $\Delta\ln(\lambda)=1/3000$ (see \S~4.1,
eq.~(\ref{eq-log-lambda-grid-increment})), it follows that we allow
$k$ to vary up and down from zero by $0.05/(1/3000)=150$.
If the minimizing value of $k$ turns out to be one of the $k$ limits $-150$ or $150$, 
then we consider the test invalid since
the true minimizing value of $k$ is almost certainly beyond the $k$ limits.
This occasionally happens and indicates the tested spectra are likely not very similar and likely cannot
be made to look similar by shifting them relative to each other in logarithmic wavelength:  if they
can, the resemblance is likely accidental.
Also we only consider DIFF2 tests valid if the good overlap region in logarithmic wavelength
(which is specified in \S~4.1) when the 
locally-normalized spectra are not relatively shifted (i.e., when $k=0$) is $\geq 0.2$ which 
corresponds to a good overlap region in relative wavelength
of $\sim 22\,$\% relative to the lower wavelength of the region.
This is the same overlap criterion for validity as used for the DIFF1 test and, as mentioned in \S~4.1,
is almost always met for spectra we consider.

    By its nature, ${\rm DIFF2}\leq{\rm DIFF1}$ for any spectrum pair.
Thus, DIFF2 has an upper limit of 2 like DIFF1 (see \S~4.1).
But it is even less likely for DIFF2 than for DIFF1 that any value will approach 2.
To do that would imply that the locally-normalized spectra were anticorrelated no matter how one shifted them
over the allowed shifting range.
Since even accidental anticorrelations of spectra in a large sample will tend to be avoided by the
shifting, we do not expect DIFF2 values that are well over 1.

    We have subjected the set of all possible spectrum pairs drawn from the sample of HDCC supernovae described in \S~5
to the DIFF2 test.
(There are 17 supernovae, 168 spectra, and 25172 valid spectrum pair tests:  
some possible pair tests are excluded  
as invalid by the validity criteria discussed above.)
% One validity criterion is the same as for DIFF1 and the other is the
% shift to the limit exclusion criterion.
The resulting distribution of DIFF2 values has
mean, standard deviation, minimum value, and maximum value of, respectively, 0.784, 0.116, 0.229, and 1.131
(where we have reported more digits than are significant to allow for numerical consistency checks).
The mean, minimum, and maximum values are, of course, 
a bit smaller than the corresponding values for DIFF1 reported in \S~4.1:
all the DIFF2 values that go into determining these distribution characteristics 
are equal to or smaller than the corresponding
DIFF1 values used to evaluate the distribution characteristics in \S~4.1. 
The standard deviation value shows that spectrum pairs significantly (i.e., $1\sigma$) 
more alike than average will have DIFF2
values of $\lesssim 0.67$.
The maximum value as expected is not well over 1.
% On the other hand, the mean 0.78 is so close to 1 as to indicate that spectrum pairs on average
% are not much alike even with our allowed shift..

    The rationale for DIFF2 is as follows.
The structure of P-Cygni line profiles in supernovae is correlated with the velocity structure
of the ejecta since that velocity structure determines the Doppler shifts of line features.
Most noticeably, weak, unblended lines in the photospheric phase tend to have their absorption features reach
a minimum at a wavelength shift corresponding to the photospheric velocity \citep[e.g.,][p.~188]{jeffery1990}.
All parts of the absorption feature have blueshifts that are usually correlated 
with the photospheric velocity in the
sense that the greater the photospheric velocity, the greater the blueshift of each part of the
absorption feature:  this is just because the whole velocity structure of the line-forming region
at any phase tends to be correlated with the photospheric velocity at that phase.
In comparing photospheric spectra for similar supernovae,
there is often some difference in the width scale of line profiles, particularly the absorption
features, attributable to some difference in the velocity structure. 
The difference may be intrinsic to the supernovae:  one supernova may just have layers that move
faster than the corresponding layers in the other supernova at the same phase.
On the other hand, the difference may just be a matter of phase since the photosphere recedes
into the ejecta as time passes because the overall optical depth of the supernova falls.
Of course, intrinsic and phase differences in velocity structure are usually both present to one degree or another.

   The eye can usually discern similar line profile patterns despite differing velocity structure. 
However, the DIFF1 test may not pick out a striking similarity in two spectra seen by the eye because 
Doppler-shift-caused offsets of line profiles between the two spectra
may lead to large contributions to the DIFF1 value.
One can imagine compensating for the different velocity structures by a wavelength varying
shift, but this seems too complex for practical use.
Since the blueshifted absorption features are often larger than the emission
features (necessarily the case if the lines are pure scattering
\citep[e.g.,][p.~189]{jeffery1990}), a general wavelength shift that tends to align the absorption features
may bring out similarities both to the eye and in the DIFF2 test (i.e., by minimizing the DIFF2 function value). 
\citet{branch2006a} showed that general shifts were useful in bringing out similarities.
We expect that the DIFF2 test shift will usually minimize the DIFF2 function value by 
tending to align blueshifted absorption features in photospheric phase spectra. 

    In the nebular phase, the emission lines tend to be symmetric about the rest-frame line-center wavelength.
If this is the case, the minimizing shift of the DIFF2 test is likely to be small or zero and
DIFF2 will tend to reduce to DIFF1.
However, if the emission peaks of one or both locally-normalized spectra
are offset from the line-center wavelengths because of some asymmetry in the ejecta,
then the DIFF2 test will compensate for that and again bring out similarities that could be missed
by the DIFF1 test.

    Because DIFF2's logarithmic wavelength shift will often compensate for Doppler shifts
caused by  differences in velocity structure, it is useful to define a corresponding 
Doppler shift velocity parameter
$v_{\rm param}$ characteristic of the difference in velocity scale between two locally-normalized spectra. 
We first preliminarily define $v_{\rm param}$ 
for a wavelength shift $\delta\lambda$ from an initial wavelength $\lambda$
using the 1st order Doppler formula (eq.~(\ref{eq-Doppler-formula}) in \S~2.2, {\it mutatis mutandis}):
thus we have
\begin{equation}
 {\delta\lambda\over\lambda}={v_{\rm param}\over c}  \,\, ,
\label{eq-velocity-param-1st}
\end{equation}
where we define positive velocities as giving redshifts.
Say the corresponding logarithmic wavelength shift is $\delta\ln(\lambda)$ and
recall we only allow logarithmic wavelength shifts of up to $0.05$.
The allowed logarithmic wavelength shifts are sufficiently small that we can then make the approximation 
\begin{equation}
\delta\ln(\lambda)\approx {\delta\lambda\over\lambda} \,\, .
\label{eq-velocity-param-2nd}
\end{equation}
We now give the final definition for $v_{\rm param}$ by combining 
equations~(\ref{eq-velocity-param-1st}) and~(\ref{eq-velocity-param-2nd}):
\begin{equation}
 v_{\rm param} = c\delta\ln(\lambda)  \,\, .
\label{eq-velocity-param}
\end{equation} 

\subsection{An Example of Two Locally-Normalized Spectra Fitted by Minimizing the DIFF2 Function Value}

    Figure~6 shows an example of two locally-normalized spectra 
with the dotted-line spectrum given the logarithmic wavelength shift that minimizes the
DIFF2 function value for
the two spectra.
The solid-line spectrum is the SN~1993J 1993 April~15 spectrum
(which comes from about 2~days before UVOIR bolometric maximum light) % redundant with below.
shown in Figure~2 (\S~3.2) and the dotted-line
spectrum is the SN~1987K 1987 August~9 spectrum \citep{filippenko1988}
(which comes from about 9~days after a red-optical maximum light). % redundant with below.
In this case, 
${\rm DIFF1}=0.642$ 
and
${\rm DIFF2}=0.612$ and the Doppler shift velocity parameter (see definition eq.~(\ref{eq-velocity-param})
in \S~4.2) corresponding to
DIFF2's logarithmic wavelength shift has absolute value $999\,{\rm km\,s^{-1}}$.
In this case, DIFF2's logarithmic wavelength shift is a blueshift when applied to the SN~1987K spectrum:
as noted above we have given this shift to the SN~1987K spectrum in Figure~6.
The change from DIFF1 to DIFF2 and the Doppler shift velocity parameter are not large in this case and, in fact,
the eye cannot detect any overall improvement in fit in going from a plot without the shift
to the plot with it (i.e., Fig.~6). 
The small change between the DIFF1 and DIFF2 and the small shift is to be expected. 
First, the spectra both come from Type~IIb supernovae.
Second, out of our current and preliminary sample of hydrogen-deficient spectra (see \S~5), 
the SN~1987K 1987 August~9 spectrum
is the closest match according to DIFF2 to the SN~1993J 1993 April~15 spectrum, except for 
other spectra from SN~1993J that come from phases close to 1993 April~15.
%% If it is a closest match, then we expect DIFF1 to be small and a small shift:  seems clear enough.
According to DIFF1, it is only the second closest non-SN-1993J match:  the 
SN~1987K 1987 August~7 spectrum is slightly closer with ${\rm DIFF1}=0.624$.
%% The difference in match between the 1987 August~9 and  1987 August~7 spectra is probably insignificant.
%% This could be reworded to be clearer, but it's past the point of no return now.

    The phase (day~$-2$) of the SN~1993J spectrum is relative to the well determined UVOIR bolometric maximum light
(see \S\S~3 and~3.2).
The phase (day~$+9$) of the SN~1987K spectrum is relative to a red-optical maximum that is estimated to
be 1987 July~31 with an uncertainty of $\pm 4$~days \citep{filippenko1988}.
Supernova SN~1993J was a particularly well-observed supernova, and hence our precise knowledge of its phase.
Supernova SN~1987K was only moderately well-observed, and hence our uncertainty about its phase and 
also the light curve from which the phase was determined.
The data available for SN~1987K is typical for the supernovae that must 
dominate any current sample of supernovae for statistical spectral analysis.
% This highlights the need for goodness-of-fit tests like DIFF1/2 that can extract information
% from only moderately well-observed or poorly observed supernovae.

    The identifications in Figure~6 are the same as in Figure~2 (\S~3.2), except for the residual telluric absorption
lines in the SN~1987K spectrum \citep{filippenko1988}.
The region from the blue side of the telluric lines to the red end of the SN~1987K spectrum were excluded from
the evaluation of DIFF1/2:  i.e., this region was not part of the 
good overlap region in logarithmic wavelength
for the spectrum pair (\S\S~4.1 and~4.2). 

\subsection{Comparing DIFF1 and DIFF2}

    Both DIFF1 and DIFF2 measure the physical similarity of line formation, and thus of supernovae.
DIFF1 puts more weight on overall physical similarity of the supernovae than DIFF2
because the DIFF2 shift 
in logarithmic wavelength
compensates somewhat for some physical distinction in the supernovae.
A spectrum pair with very small DIFF1 value may well come from 
nearly identical supernovae at nearly the same phase.
On the other hand, 
a spectrum pair with a very small DIFF2 value, but not a small DIFF1 value,
may well come from supernovae that are in some ways physically alike, but are physically
distinct in some other ways. 

    The most obvious physical distinction that the logarithmic wavelength shift 
in the DIFF2 test can compensate for
is in velocity structure:  the distinction is caused either by an intrinsic difference or a phase difference.
Another possible distinction is in viewing direction for supernovae that are asymmetric. 
Note that asymmetric supernovae even if identical would have different spectral evolution 
depending on viewing angle. 
There is, in fact, considerable evidence from supernova polarimetry and 
spectropolarimetry for asymmetry in core-collapse supernovae \citep[e.g.,][]{wheeler2000,wang2001,leonard2005}.
One important kind of asymmetry can be parameterized by a major-to-minor axis ratio for characteristic orthogonal axes 
perpendicular to the line of sight.
Spectropolarimetric observations suggest that this axis ratio can vary from 1 to 2 or more
\citep[e.g.,][]{wang2001}. 
For example, the Type~IIb supernovae SN~1993J and \object{SN 1996cb} spectropolarimetric data suggest 
an axis ratio of $\gtrsim 1.4$ for
both these supernovae \citep{wang2001}.
There are data that suggest, but with considerable uncertainty, that asymmetry increases with decreasing hydrogen
envelope mass \citep{wang2001,leonard2005}.
Another kind of asymmetry in core-collapse supernovae, that of a jet (or bipolar jets) emerging from the main part of the ejecta, 
is suggested by the spectropolarimetry of Type~Ic hypernova \object{SN 2002ap} \citep{kawabata2002,leonard2002}.
Type~Ia supernovae also show polarization in some cases:   the polarization may arise from clumps in
the ejecta \citep[e.g.,][]{leonard2005b}. 

    Because of the use of local normalization, both DIFF1 and DIFF2 tests are useful
for studying supernovae which frequently have uncertain continua.

\section{A PRELIMINARY STATISTICAL ANALYSIS OF HYDROGEN-DEFICIENT CORE-COLLAPSE SUPERNOVA SPECTRA}

    We are beginning a project of comparative analysis of the spectra of
HDCC supernovae (i.e., hydrogen-deficient core-collapse supernovae of Types IIb, Ib, Ic, and Ic hypernovae:
see \S~\ref{introduction}) that will include a statistical analysis of the spectra as well a spectral modeling
using the parameterized spectrum-synthesis code SYNOW
\citep[e.g.,][and references therein]{branch2003,branch2005}.
We note that Types IIb, Ib, and Ic seem to be a sequence of decreasing hydrogen and helium in the 
supernova ejecta.
Type~Ib supernovae have conspicuous helium lines, but not conspicuous hydrogen lines.
Type IIb supernovae have conspicuous, but weak hydrogen lines and resemble Type~Ib supernovae.
Type~Ic supernovae lack conspicuous hydrogen lines and conspicuous helium lines.
% (as mentioned in \S~\ref{introduction}).
As mentioned in the \S~\ref{introduction}, 
Type~Ic hypernovae are particularly energetic Type~Ic supernovae that in some cases are associated
with GRBs:  e.g., \object{SN 1998bw}  % verified for object macro. 
with \object{GRB 980425} \citep[e.g.,][]{galama1998}. 
All the core-collapse supernovae lack the strong Si~II~$\lambda6355$ line that characterizes
Type~Ia supernovae.
The spectral classification of supernovae is well reviewed by \citet{filippenko1997}, but note that
this paper precedes the discovery of Type~Ic hypernovae.

    We need to make clear that the HDCC supernova types are not completely distinct and the determination of
whether particular lines are conspicuous or not is sometimes ambiguous. 
For example, relatively inconspicuous hydrogen lines do seem to be present in Type~Ib supernovae
\citep[e.g.,][]{branch2002}, and so drawing the line between Type~IIb's and Type~Ib's cannot 
always be clear.
Likewise there is sometimes difficulty in distinguishing Type~Ib's from Type~Ic's and some
supernovae that cannot be clearly set in either category are often classified as Type~Ibc supernovae.   
On the other hand, the work of \citet{matheson2001} shows that the distinction between Type~Ib's and Type~Ic's
is strong enough to continue requiring the two types rather than a single merged type.
% Matheson says so in the abstract.

    To undertake the HDCC supernova analysis project, we are assembling HDCC supernova spectra
in html files in supernova directories at SUSPEND (see \S~1) under the heading {\it Supernovae by Epoch}.
The spectrum files have header information about the spectra and the supernovae they come from.
Plots of the spectra in original form and locally-normalized form are also given in the headers.
The original spectra follow the header in two-column format:  the columns being wavelength and flux. 
The locally-normalized spectra are given in the dif subdirectories of the supernova directories.
A list of the spectra are currently in html files (including those spectra from non-HDCC
supernovae) can be found under the heading {\it Lists} at SUSPEND.

    At present, we are at a very early stage in the project and have not assembled many of the
spectra available for HDCC supernovae nor determined the best
way to proceed with our statistical analysis.
However, to illustrate the use of DIFF1 and DIFF2, we present some preliminary results. 

\subsection{Table 1 and Tables at SUSPEND}

    Table~\ref{table1} gives some statistics on the HDCC supernovae and spectra we have assembled so far in SUSPEND
(see \S~1).
As one can see from the table, there are currently 17 supernovae and 168 spectra in the sample.
The supernovae include the prototype Type~Ib supernovae
\object{SN 1983N} and \object{SN 1984}L \citep{harkness1987}
and the well-observed Type~Ic's \object{SN 1987M} \citep[][]{filippenko1990} and     % 83I and 83V are Ic 
prototypes.
\object{SN 1994I} \citep[e.g.,][]{clocchiatti1996}.
The peculiar Type~Ibc \object{SN 2005bf} \citep[e.g.,][]{folatelli2006,parrent2007} is included as well:
this supernova had two observed maxima, had a rise time of about 40~days to the second 
and main optical maximum (typical
Type Ib/c rise times are of order 20~days), and was unusually luminous.   % See Folatelli 1042 & 1049
(Supernovae also have a near-explosion maximum light that is almost never observed, and so
usually unmentioned.
This is not the maximum light we refer to here as the first maximum light of SN~2005bf.)
SN~2005bf also made a transition from Type~Ic to Type~Ib behavior in the period from about day~$-30$
to about day~$+20$ relative to the (main) optical maximum \citep[e.g.,][]{folatelli2006}: % p. 1043--1044
because of this transition, SN~2005bf is classified as a Type~Ibc. 
% (Type~Ibc supernovae are those that are not clearly distinguishable into Type~Ib's or Type~Ic's.) % Said above.
Only two Type~IIb supernovae have been included so far:  the already discussed
prototype SN~1987K (\S~4.3;  \citealt{filippenko1988}) and the very well-observed
SN~1993J (\S~3 and \S~4.3; e.g., \citealt{jeffery1994,richmond1994,baron1995,clocchiatti1995,houck1996,
fransson2005}).
Three Type~Ic hypernovae are included:  \object{SN 1997ef} \citep[e.g.,][]{mazzali2000},
SN~1998bw \citep[e.g.,][]{galama1998,iwamoto1998,patat2001}, and
SN~2002ap \citep[e.g.,][]{kawabata2002,leonard2002,foley2003}.
Also included are less famous HDCC supernovae observed by \citet{matheson2001} and others.   % 90aa may be from others.

     In the calculations described below, we included only one spectrum for each phase period (which we
usually set to being 1 day) from each supernova.
The other spectra for that phase period are redundant since they should be nearly identical
to the included spectrum aside from observational error.
The spectrum we included is the one we deem to be best:  this is usually because of broader wavelength
coverage. 

     In an automated fashion, we have calculated DIFF1 and DIFF2 for all supernova
spectrum pairs in the sample:  these spectrum pair DIFF1/2 values can be found tabulated at SUSPEND under
the heading {\it Lists}.
For each spectrum, there is a table listing all other spectra in order of increasing
DIFF1/2 values relative to it.
These tables are updated as more supernovae and spectra are added to the sample.
We have excluded from the analysis below and in \S~5.2 those
DIFF1 and DIFF2 values deemed invalid by the rules discussed
in \S\S~4.1 and~4.2.  

      As well as ordering spectra in order of likeness using DIFF1/2 values, 
we want some way to order supernovae in order of likeness.
As a preliminary method, for any two supernovae we calculate all (valid) DIFF1/2 values between
spectrum pairs with one of the pair coming from one supernova and the other from the
other supernova:  i.e., we calculate the non-self spectrum pair DIFF1/2 values. 
The smallest of these DIFF1/2 values we call the supernova pair DIFF1/2 value.

    The rationale for the definition for supernova pair DIFF1/2 value is as follows.
The spectral phase coverage for all supernovae is incomplete (although for some the coverage is very good)
and rarely are both supernovae covered at a common phase to within a day:  often
coverage at even a nearly common phase is lacking.
Also phase measured from a time zero set at maximum light (which is how we define phase following a common convention) 
is itself uncertain.
Ideally, one would like to use an accurate UVOIR bolometric maximum light as time zero for supernova phase.
We do that when an accurate UVOIR bolometric maximum light is available (which is rarely).
Usually, however, we must use a substitute which is often some optical broad-band maximum light.
These broad-band maxima usually happen within a few days of UVOIR bolometric maximum light and have some uncertainty. 
In some cases, all we can use as a substitute is what can be loosely called optical  
maximum light which is a fiducial time within a few days of when 
exact UVOIR bolometric maximum light probably occurs. 
And in some of these cases, the estimate of optical maximum light is just the day of discovery:  supernovae tend to be
discovered near or at optical maximum light since that is when they are most readily discoverable because,
of course, they are brightest or nearly brightest in all optical bands at optical maximum light.  
(Throughout the rest of this section we usually do not specify the kind of maximum light 
we use to set the phase zero point.
We usually just refer to all kinds of maximum light as maximum light without qualification.)
By choosing the smallest DIFF1/2 value out of all the non-self spectrum pair values 
for the supernova pair DIFF1/2 value, we hope to have partially avoided the problems 
of incomplete spectral phase coverage
and inaccurate phase measured from UVOIR bolometric maximum light.
We assume that the intrinsic likeness of each supernova pair is well measured by their
closest observed spectral approach to each other:  i.e., what we call
the supernova pair DIFF1/2 value. 
Obviously, because of the incomplete phase coverage, the closest observed spectrum approach
may only give a lower limit on likeness. 
For two supernovae that are actually much alike with at least one having good phase coverage, 
the closest spectrum approach is probably for the spectrum
pair that has the least relative phase difference where in this case phase is measured relative
to explosion. 
(As supernovae evolve from time zero at explosion, 
their relative rate of spectral evolution tends to be more constant
than their absolute rate of spectral evolution.) 
For two supernovae that are not much alike or that both have poor phase coverage, 
the closest spectrum approach may not occur for the spectrum
pair that has the least or even small relative phase difference. 

     There are, in fact, cases where two supernovae 
have a relatively small supernova pair DIFF1/2 value and yet are in some respects quite different.
For example, the day~$+4$ (relative to $B$ maximum) spectrum of 
Type~Ibc \object{SN 1999ex} \citep{hamuy2002,stritzinger2002}
is very similar to the day~$-20$ (relative to UVOIR bolometric maximum light)
spectrum of Type~Ibc SN~2005bf \citep{folatelli2006,parrent2007}.
Figure~10 of \citet{parrent2007} shows overlapping locally-normalized versions of these
spectra:  to the eye their fit is very close.
(See also Fig.~5 of \citealt{folatelli2006}.)
The supernova pair DIFF1 and DIFF2 values are, respectively, 0.447 and 0.442:  these are quite
small DIFF1/2 values for spectra from different supernovae as our experience shows.
Although SN~1999ex and SN~2005bf are both Type~Ibc supernovae, they are quite different
in some respects:  e.g., rise time to optical maximum:  probably of order 20~days for
SN~1999ex \citep{richardson2006}
and about 40~days to the second
and main optical maximum, as mentioned above, for SN~2005bf.
Also the two closely matching spectra come from very different phases
which would not be true if the supernovae were really very much alike:  the most closely matching
spectra would then tend to come from nearly the same phase.
Thus, it is clear that our assumption of intrinsic supernova likeness being well measured
by the supernova DIFF1/2 pair is not always valid.
However, since our analysis here is preliminary, we leave the search for a better 
measure of supernova likeness to future work {\it sine die}.

        The tables giving the spectrum pair DIFF1/2 values at SUSPEND (which altogether
are quite lengthy) are preceded by tables of supernova pair DIFF1/2 values. 
For each supernova there is a table listing all other supernovae in order of increasing 
supernova pair DIFF1/2 values relative it.

\subsection{Table 2 and the Standard HDCC Supernova Types}

       Since this paper is just a beginning in our project of statistical analysis
of HDCC supernova spectra, we will not fully discuss the significance of either
the tabulated supernova or spectrum pair DIFF1/2 values at SUSPEND (see \S~5.1).
But as a preliminary investigation, we consider if the 
classification of HDCC supernovae into Type~IIb's,  Type Ib's, Type~Ic's,
and Type~Ic hypernovae is discoverable from supernova pair DIFF1/2 values.
(Supernovae classified as Type~Ibc are grouped with the Type~Ib's.
There are only two of these in the current sample:
SN~1999ex \citep{hamuy2002,stritzinger2002} 
and
SN~2005bf \citep[e.g.,][]{folatelli2006,parrent2007}.)
A~priori, it is not clear that the HDCC supernova types will be picked out by
an analysis with DIFF1/2.
The Types IIb, Ib, and Ic, and, in part, the Type~Ic hypernova are characterized by particular lines.
DIFF1/2 measures overall differences between spectra. 

     For the preliminary investigation, we have calculated the mean supernova pair DIFF1/2 values between supernovae
of different types (the cross-type means) and 
the mean supernova pair DIFF1/2 values between supernovae of the same type
(the self-type means). 
We have also calculated the estimated standard deviations of the distributions of the
supernova pair DIFF1/2 values about the type means 
%  (when there is more than one DIFF1/2 value available for the type-mean calculation)   % Said below.
using the ordinary standard deviation formula with the correction for using the mean of the sample
\citep[e.g.,][p.~19]{bevington1969}.
The standard deviation for the Type~IIb self-type mean is assigned a zero value
since there are only two Type~IIb's in the current sample, and thus only one supernova pair DIFF1/2 value:  
the standard deviation of the sample is zero;  that of the distribution is unknown.

     Since HDCC supernovae tend to look increasingly alike
as time from explosion (i.e., phase) increases, we have repeated the
DIFF1/2 calculations including only supernova pair DIFF1/2 values obtained from spectra
that are both from phases less than 10~days past maximum light.
The supernova pair DIFF1/2 type means obtained in this case are phase-restricted.
We were interested in seeing if phase-restricted approach would better distinguish HDCC supernovae
into the standard types than the phase unrestricted approach. 
Note that phase-unrestricted supernova pair DIFF1/2 values will, of course, 
always be smaller than or equal to their
counterpart phase-restricted supernova pair DIFF1/2 values.
(The last remark assumes that counterpart phase-restricted supernova pair DIFF1/2 values exist.
They may not if one or both of the supernovae in the pair lack any spectra for the period allowed
by the phase restriction.)

     The results of all the calculations are given in Table~2 in the form of a type DIFF1/2 correlation matrix
consisting of self-type means (the diagonal elements) and cross-type means (the off-diagonal elements):
phase-unrestricted DIFF1 values are in Table~2(a), phase-unrestricted DIFF2 values are Table~2(b), phase-restricted DIFF1 values
are in Table~2(c), and phase-restricted DIFF2 values are in Table~2(d).
The matrix is, of course, symmetric.
% We will denote elements of the matrix like so:   (Ib,Ic) for the mean DIFF1/2 between Type~Ib's and Type~Ic's,
% etc.  % I apparently never use this notation below.

       Several remarks can be made about the results in Table~2.
First, none of the mean values are very small:  the smallest is $0.455$ which
is the Type~Ic hypernova DIFF2 self-type mean in Table~2(b). 
An examination of the table of the phase-unrestricted supernova pair DIFF1/2 values at SUSPEND 
shows that none are less $0.4$:
the smallest DIFF1 is $0.418$ for the pair of Type~Ib \object{SN 1988L} and Type~Ic \object{SN 1990B};
the smallest DIFF2 is $0.409$ for the same pair of supernovae and obtained using the same spectrum pair. 
The fact that these smallest DIFF1/2 values are for supernovae of different assigned types shows
that supernova pair DIFF1/2 values are not perfect in discriminating Type~Ib and Type~Ic supernovae.  
Some of the spectrum pair DIFF1/2 values are significantly smaller (some below 0.3 going down to
minimum values of 0.244 for DIFF1 (see \S~4.1) and 0.229 for DIFF2 (see \S~4.2)), but 
those are for spectra from the same supernova usually taken close together in phase. 
The upper limit on the phase-unrestricted supernova pair DIFF1 values is 0.886 and on the 
phase-unrestricted supernova pair DIFF2 values is 0.809.
These upper limits were both
for the supernova pair of Type~Ic \object{SN 1990aa} and Type~Ib \object{SN 1999di}, but not using 
the same spectrum pair.
The supernovae of this pair are not much alike in the phases for which spectra are available. 

      Second, a mean DIFF2 value is always less than or equal to the corresponding mean DIFF1 value 
as it must be according to our formulae (\S~4.1, eq.~(12) and \S~4.2, eq.~(17)), but always by less
than the standard deviation of the mean DIFF1 value. 
We have to conclude that DIFF2 is not finding great similarities that are being totally
missed by DIFF1.
% On the hand, in some cases the DIFF2 standard deviations are significantly smaller   % A closer look shows no clear case for this
% than the corresponding DIFF1 standard deviations.                                    % conclusion.
% This suggests that DIFF2 is better at gauging the actual similarity level by compensating
% for phase differences.

      Third, the phase-restricted mean DIFF1/2 values 
(in Table~2(c) and Table~2(d)) are all larger than the corresponding
phase-unrestricted mean DIFF1/2 values (in Table~2(a) and~(b)).
This is understandable for two causes.
First, HDCC supernovae are qualitatively understood to be more diverse at earlier times, and thus
the phase-restricted DIFF1/2 values should tend to be larger.
Second, the incompleteness of the phase coverage of a supernova may be greatly increased 
by phase restriction.
Phase restriction reduces the chances of finding supernova pair DIFF1/2 values for the 
aforementioned supernova with another
supernova with pair spectra from nearly the same phase
when the spectra would tend to be most alike if the pair supernovae were alike. 
The effects of the two causes could be sorted out in a more detailed analysis.
It is possible for the mean DIFF1/2 values to be reduced by phase restriction.
This would happen if phase restriction eliminated a supernova altogether from those going
into calculation of the means and that supernova was sufficiently remote from other supernovae
to increase phase-unrestricted mean DIFF1/2 values when included.
This actually happened in an earlier calculation for the Type~Ic self-type means.
The phase-restriction eliminates Type~Ic SN~1990aa for which no spectra from before 10~days 
after maximum light are currently available to us. 
SN~1990aa may be something of an outlier among Type~Ic's, and so its inclusion seems to increase the
self-type mean DIFF1/2 values.
When the sample of Type~Ic's was changed this reduction on phase restriction disappeared. 
     
     Fourth, the self-type means in any row are the smallest means in that row in all cases. 
This shows that a DIFF1/2 analysis can probably identify the standard HDCC supernova types on average.
If one is given spectra from a new HDCC supernova of unknown type and uses those spectra to find 
mean supernova pair DIFF1/2 values with  % "spectra from the"  :  we are talking of means with samples.
samples of the standard types, 
the smallest mean supernova pair
DIFF1/2 value will probably be with the sample of the type to which the new HDCC supernova belongs. 

     In the rows for Type~IIb's and Type~Ic hypernovae, the self-type means are  
smaller than the respective row cross-type means by $\gtrsim 1.5\sigma_{\rm larger}$ in 
all cases and by $\gtrsim 2\sigma_{\rm larger}$ in all but two cases.
(The $\sigma_{\rm larger}$ is the larger 
of the standard deviations for the two means being compared.)
The two cases where difference is less than $2\sigma_{\rm larger}$ are for the
DIFF1 and DIFF2 phase-unrestricted Type~Ic hypernovae cross-type with the Type~Ic's.  
Overall, Type~IIb and Type~Ic hypernova are found to be quite distinct from 
other types.

    On the other hand, in the phase-unrestricted rows for Type~Ib/Ibc's 
and phase-unrestricted and phase-restricted rows for Type~Ic's the self-type means are
smaller than the respective row cross-type means by $\lesssim\sigma_{\rm larger}$.
For the phase-restricted cases, the Type~Ib/Ibc self-type means in are some cases separated by
more than $\sigma_{\rm larger}$ from the Type~Ib/Ibc row cross-type means.
Overall, one must conclude that Type~Ib/Ibc and Type~Ic are not as distinct from
other types as are Type~IIb and Type~Ic hypernova.  
This means that Type~Ib/Ibc and Type~Ic are more heterogeneous and overlap more in behavior
with other types than Type~IIb and Type~Ic hypernova.

    Recall that we introduced phase restriction to see if it
would better distinguish HDCC supernovae into the standard types than the phase unrestricted approach. 
From the results discussed in the last two paragraphs,
there is some improvement in distinguishing the types.
However, this improvement is rather modest for the particular phase restriction we used:  i.e.,
restricting spectra used to those from before 10~days after maximum light.
Maybe other phase restrictions would give greater improvement.
This could be investigated in future work.

%     We note from the above discussion that Type~Ib/Ibc's and Type~Ic's seem to be more heterogeneous
% types than Type~IIb's and Type~Ic hypernovae and overlap more in behavior with other types.
%%% Now redundant with the above.

    To finish our preliminary analysis, 
we note that the Type~IIb's       % 7 for 87l and 38 for 93J
and Type~Ic hypernovae   % 25 for 97ef and 17 for 98bw and 10 for 02ap.
included in the sample are all relatively well-observed
supernovae, and so they suffer relatively little from lack of phase coverage.
Some of the Type~Ib/Ibc's and Type~Ic's in the sample have poor or very poor phase coverage.
(See the list of supernova spectra in html format under the heading {\it Lists} at 
SUSPEND for the spectra currently available for individual supernovae and the spectrum phases.)
The supernova pair DIFF1/2 values for supernovae where both supernovae have poor phase coverage
could be large even if the supernovae were very similar simply because the procedure
for determining the supernova pair DIFF1/2 values may not be able to find spectra from nearly
the same phase. 
In future work, we will try to remedy this deficiency in the procedure perhaps by
giving lower weight in calculating mean DIFF1/2 values to supernova pair DIFF1/2 values from supernova pairs where
both members of the pair have poor phase coverage or by using some quite different
definition of supernova pair DIFF1/2 value.

\section{CONCLUSIONS AND DISCUSSION}

%\subsection{Conclusions}

    We have developed two tests DIFF1 and DIFF2 for measuring
goodness-of-fit between two supernova spectra (see \S~4.1, eq.~(12) and~\S~4.2, eq.~(17) for the formulae).
The tests rely on local normalization (\S~2) which eliminates the uncertainty
in the spectrum continuum, except for contamination from extraneous sources (\S~1).
Local normalization also eliminates real information stored in continuum shape.
However, a basic premise of this paper is that line pattern is a much better signature
of intrinsic supernova behavior than continuum shape, and so eliminating 
continuum shape information is not too important for spectrum comparisons (\S~1). 

    We have presented some examples of locally-normalized spectra for
SN~1993J and given some analysis of the spectra (\S~3).
The UV parts of two of the SN~1993J spectra are hitherto unpublished {\it HST} spectra.
In \S~3.5, we have shown that 
local normalization is sufficiently continuum-independent to lead to valid
results in comparing locally normalized spectra. 
One must, however, apply the same local normalization procedure to all original spectra
in order to obtain locally-normalized spectra that can be validly compared.

    As an example of the use of DIFF1/2, we have used them in a preliminary
statistical analysis of the spectra of HDCC (hydrogen-deficient core-collapse) supernovae (\S~5).
The analysis shows that standard HDCC supernova types (IIb, Ib/Ibc, Ic, 
and Ic~hypernova) do form distinct groups when 
compared using DIFF1/2 (\S~5.2).   % See the ``fourth'' point of the remarks in section 5.2.
This analysis is preliminary since many available HDCC supernovae and HDCC supernova spectra 
are not included in the analysis sample.
Also many improvements in our statistical procedure are possible.  

    Tables of all spectrum and supernova pair DIFF1/2 values
for our sample of HDCC supernovae are
available at the SUSPEND database (see the footnote to the abstract for the URL) under
the heading {\it Lists}.
The spectra we have used are also online at SUSPEND in two-column format under the heading
{\it Supernovae by Epoch} in supernova directories in html files along with figures
of the original and locally-normalized spectra.
The locally-normalized versions of the spectra
themselves can be found in the dif subdirectories of the supernova directories. 

     Although hundreds of supernovae are now being discovered per year
(e.g., 367 in 2005 and 526 in 2006 \citep{cbat2007} which gives discovery
rates of more than 1 supernova per day),
 most of these are remote and are relatively poorly observed.    % up to 2005nc gives 367.
Their main use is as cosmological distance indicators and for the determination of supernova rates.
New well-observed supernovae accumulate slowly with only a few per year. 
These new well-observed ones and the past well-observed ones are only relatively
well-observed in most cases.
Phases are missing, calibrations imperfect, and frequently the reddening correction
is very uncertain.
Thus, for the foreseeable future, statistical analyses of supernova spectra will have
to rely on heterogeneous data sets for relatively few well-observed supernovae.
We believe that DIFF1 and DIFF2 (which eliminate the need for accurate
continuum shape and that can, in the case of DIFF2, compensate somewhat for varying phase, velocity structure,
and asymmetry (\S~4.4)) will be useful tools in statistical analyses of available spectra.

    Another, and not-distinct, use for DIFF1/2 is to find for new and, perhaps, 
not-well-observed supernovae, well-observed supernovae that are their near-twins. 
Then insofar as the well-observed supernovae are understood, the new supernovae will
be understood.

     DIFF1/2 should also be useful in synthetic spectrum analysis of supernova spectra
because, again, it eliminates the 
% need for dealing with the   % This is unneeded and I don't use it elsewhere like at the top of his section.
uncertain continuum of observed supernovae.  %  at least to high accuracy. % An unclear phrase.
Certainly fitting the true continuum is one of the goals and guides in achieving a good
synthetic spectrum fit to the observations.
But when the fitting exercise gives a good fit to an incorrect continuum, then 
the fitting exercise becomes misleading.
A good fit to the lines with highly realistic radiative transfer
and a realistic hydrodynamic model should yield a good
continuum and that should allow one to correct the observed continuum for 
errors in dereddening and calibration.

     The future of synthetic spectrum modeling of supernovae may well be the calculation
of time-dependent radiative transfer for a large collection of realistic hydrodynamic
models that span many possible explosion outcomes.
As time passes more models will be included in the collection and the realism of
the models and the radiative transfer will be improved. 
Eventually, almost all old and new supernovae could find a near-twin in the collection using
some test like DIFF1/2 and will then be well understood.
This ideal situation is still far off, but there is work heading toward it
\citep[e.g.,][for Type~Ia supernovae]{woosley2007}.    

     In future work, we plan to use DIFF1/2 in fuller statistical analyses of
HDCC supernova spectra than that presented in this paper.
These analyses will also include spectral modeling using the  
parameterized spectrum-synthesis code SYNOW 
\citep[e.g.,][and references therein]{branch2003,branch2005}.
Use of DIFF1/2 for other supernova types is also envisaged for future work.
 
\acknowledgments
We would like to thank all the observers who obtained the spectra used in this paper,
Peter Challis (who prepared the two nebular-phase {\it HST} spectra for us some time ago)
and St\'ephane Blondin for information about some of the SN~1993J spectra we present,
and the referee for his/her comments.
Support for this work has been provided by NASA grant NAG5-3505
and
the Homer L. Dodge Department of Physics \&~Astronomy of the University of Oklahoma.

%\appendix
% appendix a
%\section{Appendix A}
% appendix b   No need for another \appendix.  That just gives another Appendix A.
%\section{Appendix B}

% References

\clearpage

%% Use the figure environment and \plotone or \plottwo to include 
%% figures and captions in your electronic submission.

\begin{figure}
%\plotone{./f1.eps}
\includegraphics[height=.9\textheight,angle=0]{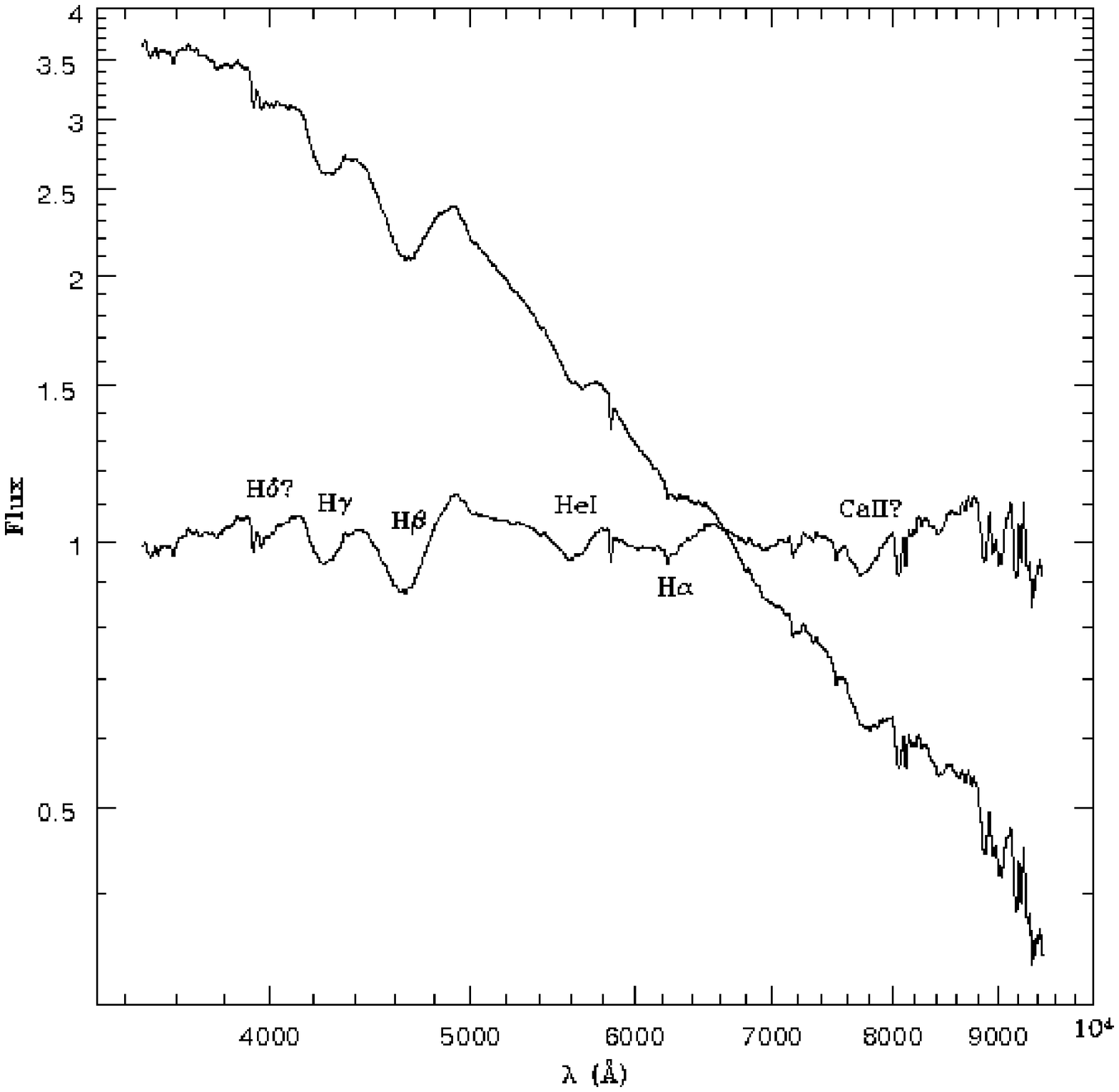}
\caption{The spectrum of Type~IIb SN~1993J in the $f_{\lambda}$ and locally-normalized representations from
1993 March 31 which is about 16~days before UVOIR bolometric maximum light and about 4~days after explosion.
The locally-normalized spectra in this and in other figures are obviously the ones with continuum
level of about 1.
\label{fig1}}
\end{figure}

\begin{figure}
%\plotone{./f2.eps}
\includegraphics[height=.9\textheight,angle=0]{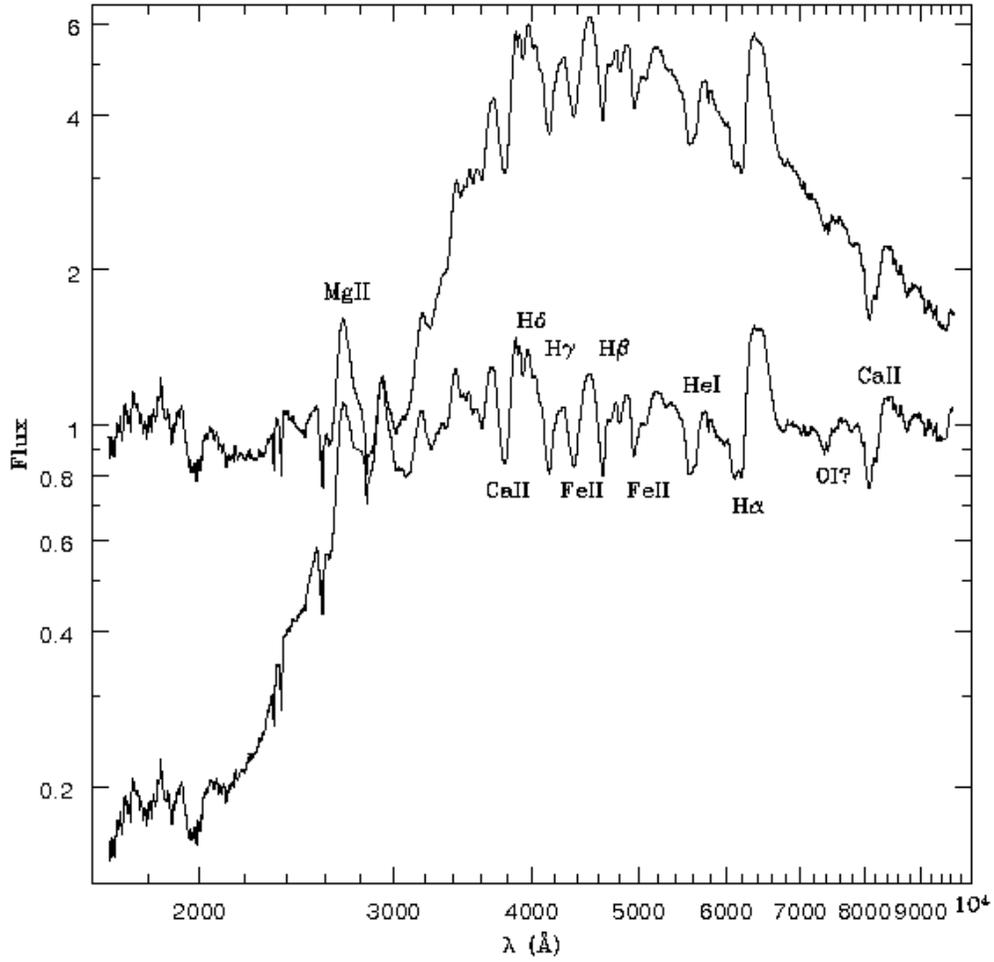}
\caption{
The spectrum of Type~IIb SN~1993J in the $f_{\lambda}$ and locally-normalized representations from
1993 April~15 which is about 2~days before UVOIR bolometric maximum light and 18~days after explosion.
From $3240\,$\AA\ blueward, the spectrum is an {\it HST} spectrum.
\label{fig2}}
\end{figure}

\begin{figure}
%\plotone{./f3.eps}
\includegraphics[height=.9\textheight,angle=0]{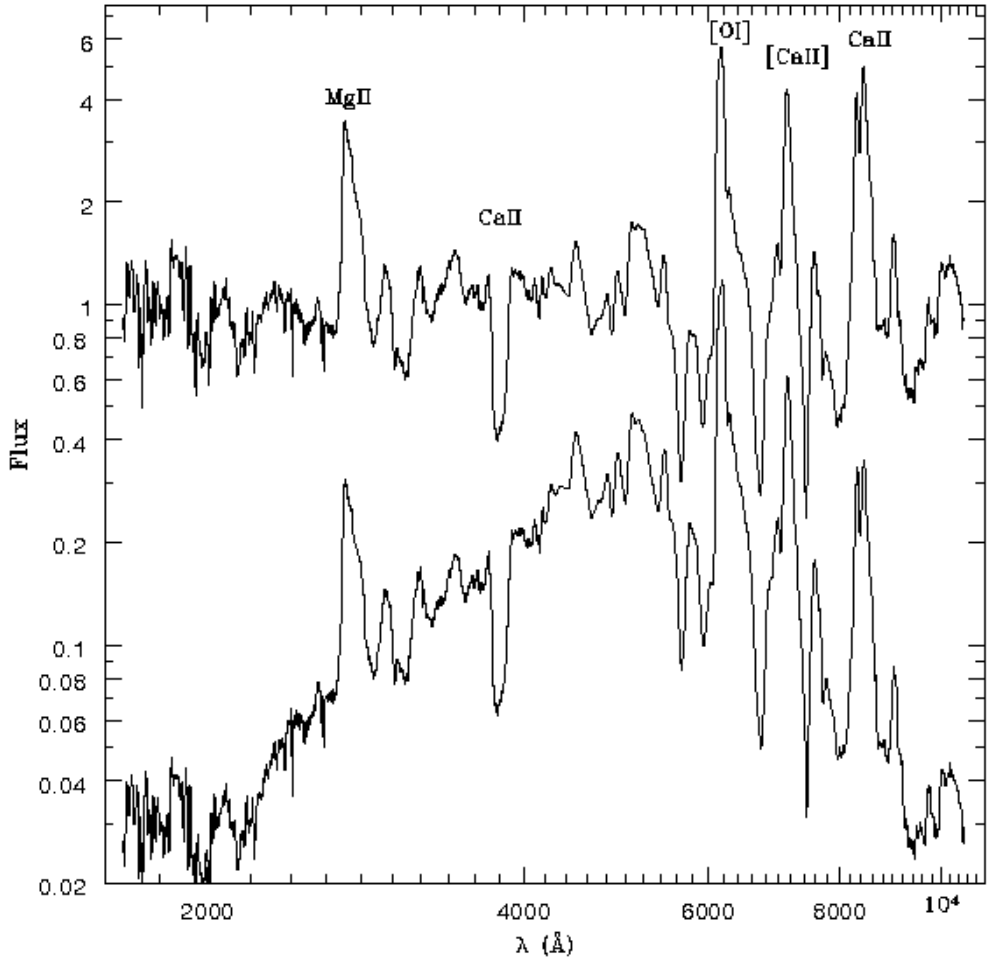}
\caption{
The spectrum of Type~IIb SN~1993J in the $f_{\lambda}$ and locally-normalized representations from
1993 September 15.5 (an averaged date)
which is about 151~days after UVOIR bolometric maximum light and 171~days after explosion.
From $4253.6\,$\AA\ blueward, the spectrum is an {\it HST} spectrum (from 1993 September 17).
\label{fig3}}
\end{figure}

\begin{figure}
%\plotone{./f4.eps}
\includegraphics[height=.9\textheight,angle=0]{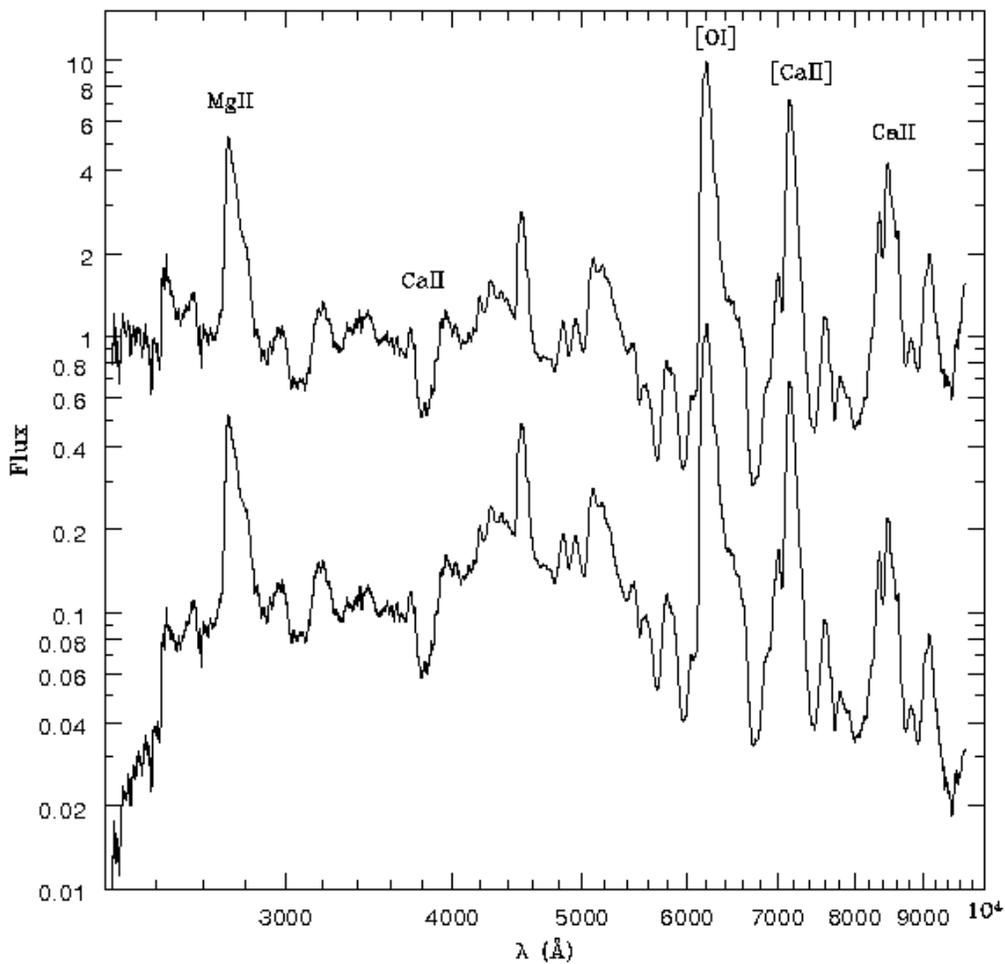}
\caption{
The spectrum of Type~IIb SN~1993J in the $f_{\lambda}$ and locally-normalized representations from
1993 November 15.5 (an averaged date) which is about 212~days after UVOIR bolometric maximum light 
and 232~days after explosion. 
From $4200\,$\AA\ blueward, the spectrum is an {\it HST} spectrum (from 1993 November 14).
The original {\it HST} spectrum blueward of $\sim 2425\,$\AA\ seems to decline in a manner too steep to be physically 
real and we do not trust it nor the locally-normalized spectrum there.
\label{fig4}}
\end{figure}

\begin{figure}
%\plotone{./f5.eps}
\includegraphics[height=.9\textheight,angle=0]{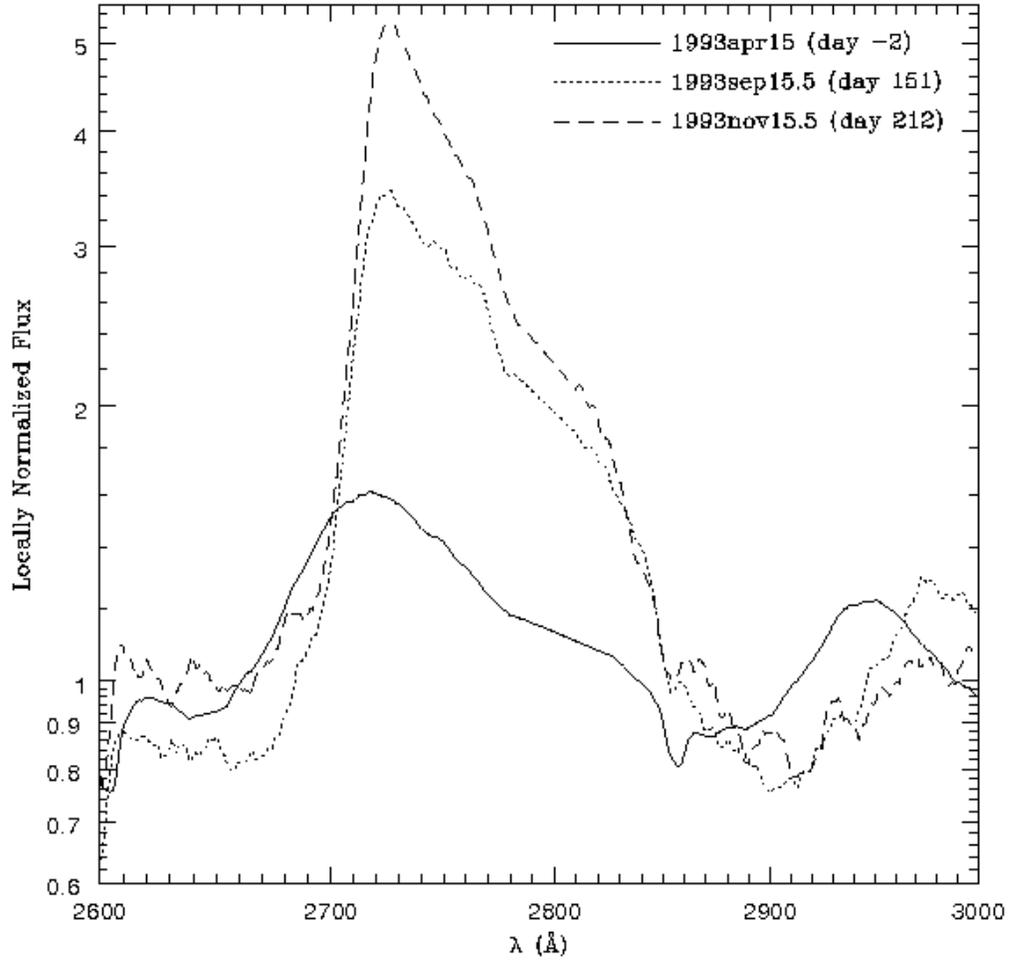}
\caption{
The evolution of the blueshifted Mg~II $\lambda2797.9$ emission line in the 
locally-normalized spectra of SN~1993J.
The phases given in parentheses in the figure are relative to the UVOIR bolometric maximum light
on 1993~April~17.
The horizontal axis is logarithmic, but this is hard to notice given
the small wavelength range covered.
\label{fig5}}
\end{figure}

\begin{figure}
%\plotone{./f6.eps}
\includegraphics[height=.8\textheight,angle=0]{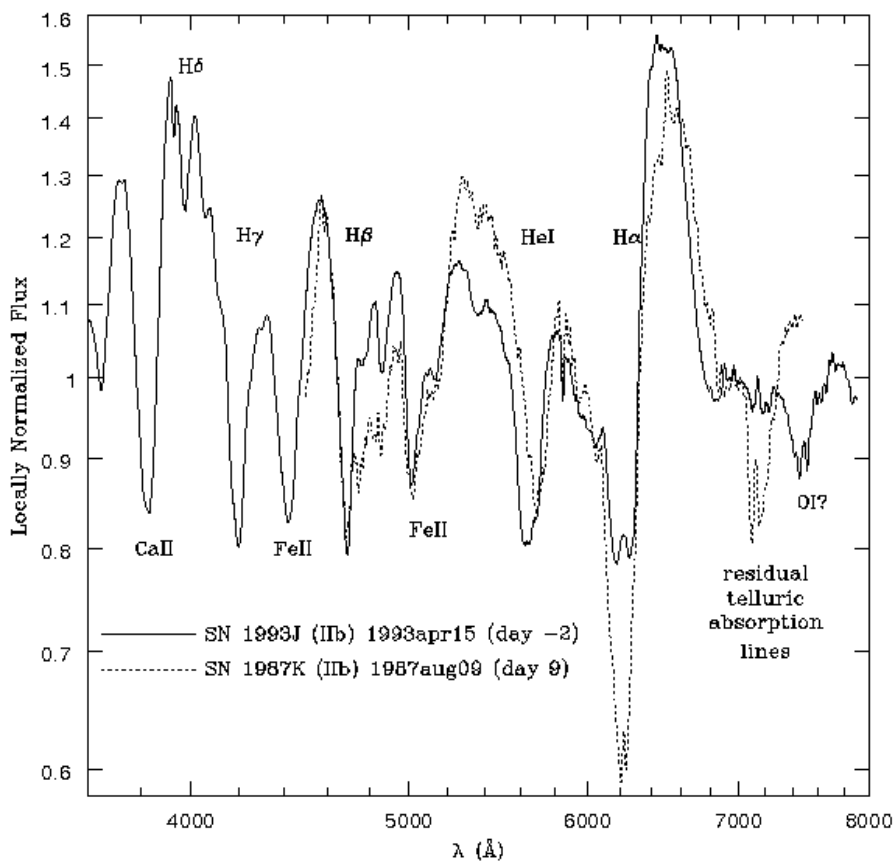}
\caption{
The locally-normalized spectra of Type~IIb supernovae SN~1993J from 1993 April~15 (about 2~days before
UVOIR bolometric maximum light) and SN~1987K from 1987 August 9 (about 9~days after a 
red-optical maximum light).
The SN~1987K spectrum has been blueshifted using Doppler shift velocity parameter
absolute value $999\,{\rm km\,s^{-1}}$ 
to minimize the DIFF2 function value for the spectrum pair.
For the spectrum pair, ${\rm DIFF1}=0.642$ and  ${\rm DIFF2}=0.612$.
The spectra were both initially corrected for host galaxy heliocentric velocity:
from Leda, the mean host galaxy heliocentric velocities are
$-39\pm2\,{\rm km\,s^{-1}}$ (SN~1993J) 
and
$799\pm2\,{\rm km\,s^{-1}}$ (SN~1987K) \citep{paturel2003}.
\label{fig6}}
\end{figure}

\clearpage

\begin{deluxetable}{lrrr}
\tabletypesize{\scriptsize}
\tablecaption{Current Sample in SUSPEND of HDCC Supernovae and Spectra\label{table1}}
\tablewidth{0pt}   % this sets the default width
%\rotate
%\tablewidth{2.truein}
\tablehead{
\colhead{Type}
&\colhead{Number of Supernovae}
&\colhead{Number of Spectra}  \\
}
\startdata
All HDCC Types  &    17 &        168 \\
Type IIb    &         2 &         39 \\
Type Ib/Ibc &         8 &         45 \\
Type Ic     &         4 &         34 \\
Type Ic hyp &         3 &         50 \\
\enddata

%\tablenotetext{a}{All types of HDCC supernovae is what is meant.}

\tablenotetext{a}{
Type~Ibc supernovae are those supernovae which are not clearly distinguishable into Type~Ib's or Type~Ic's.
We group Type~Ib's and Type~Ibc's together.
There are only two Type~Ibc' in the sample:  
SN~1999ex \citep{hamuy2002,stritzinger2002} and
SN~2005bf \citep[e.g.,][]{folatelli2006,parrent2007}.
Type~Ic~hyp is short for Type~Ic~hypernovae.
}

%\tablecomments{
%}
\end{deluxetable}

\clearpage

%\begin{deluxetable}{lrrrrrrrr}
\begin{deluxetable}{lcccccccc}
\tabletypesize{\scriptsize}
\tablecaption{HDCC Supernova Type DIFF1/2 Correlation Matrix\label{table2}}
\tablewidth{0pt}   % this sets the default width
%\tablewidth{300pt}   % this sets the default width
%\rotate
%\tablewidth{2.truein}
\tablehead{
\colhead{Type$\backslash$Type}
&\colhead{Type~IIb}
&
&\colhead{Type~Ib/Ibc\tablenotemark{a}}
&
&\colhead{Type~Ic}
&
&\colhead{Type~Ic~hyp\tablenotemark{b}} 
%&\colhead{Type~Ic~hypernova\tablenotemark{b}} 
&  \\
}
\startdata
&
Mean DIFF1/2   & St.Dev.  &
Mean DIFF1/2   & St.Dev.  &
Mean DIFF1/2   & St.Dev.  &
Mean DIFF1/2   & St.Dev.  \\ 
\cutinhead{(a) DIFF1:  no phase restriction}
Type IIb      &       0.507 &       0.000   &       0.674 &       0.077   &       0.662 &       0.068   &       0.632 &       0.052 \\
Type Ib/Ibc   &       0.674 &       0.077   &       0.615 &       0.097   &       0.646 &       0.113   &       0.712 &       0.081 \\
Type Ic       &       0.662 &       0.068   &       0.646 &       0.113   &       0.578 &       0.118   &       0.614 &       0.085 \\
Type Ic hyp   &       0.632 &       0.052   &       0.712 &       0.081   &       0.614 &       0.085   &       0.499 &       0.068 \\
\cutinhead{(b) DIFF2:  no phase restriction}
Type IIb      &       0.507 &       0.000   &       0.644 &       0.071   &       0.618 &       0.038   &       0.601 &       0.042 \\
Type Ib/Ibc   &       0.644 &       0.071   &       0.573 &       0.094   &       0.610 &       0.105   &       0.670 &       0.056 \\
Type Ic       &       0.618 &       0.038   &       0.610 &       0.105   &       0.547 &       0.120   &       0.579 &       0.071 \\
Type Ic hyp   &       0.601 &       0.042   &       0.670 &       0.056   &       0.579 &       0.071   &       0.455 &       0.020 \\
\cutinhead{(c) DIFF1:  restricted to pre-10-days past maximum light}
Type IIb      &       0.507 &       0.000   &       0.730 &       0.064   &       0.699 &       0.066   &       0.735 &       0.086 \\
Type Ib/Ibc   &       0.730 &       0.064   &       0.633 &       0.085   &       0.654 &       0.082   &       0.855 &       0.128 \\
Type Ic       &       0.699 &       0.066   &       0.654 &       0.082   &       0.597 &       0.160   &       0.754 &       0.135 \\
Type Ic hyp   &       0.735 &       0.086   &       0.855 &       0.128   &       0.754 &       0.135   &       0.504 &       0.072 \\
\cutinhead{(d) DIFF2:  restricted to pre-10-days past maximum light}
Type IIb      &       0.507 &       0.000   &       0.696 &       0.053   &       0.647 &       0.038   &       0.681 &       0.069 \\
Type Ib/Ibc   &       0.696 &       0.053   &       0.593 &       0.076   &       0.623 &       0.071   &       0.763 &       0.074 \\
Type Ic       &       0.647 &       0.038   &       0.623 &       0.071   &       0.557 &       0.130   &       0.652 &       0.083 \\
Type Ic hyp   &       0.681 &       0.069   &       0.763 &       0.074   &       0.652 &       0.083   &       0.485 &       0.064 \\
\enddata

\tablenotetext{a}{Type~Ibc supernovae are those supernovae which are not clearly distinguishable into Type~Ib's or Type~Ic's.
We group the Type~Ib's and Type~Ibc's together.
There are only two Type~Ibc's in the sample:
SN~1999ex \citep{hamuy2002,stritzinger2002} and
SN~2005bf \citep[e.g.,][]{folatelli2006,parrent2007}.
}

\tablenotetext{b}{Type~Ic~hyp is short for Type~Ic~hypernovae.}

\tablecomments{This table gives the mean DIFF1/2 values between HDCC supernova types in a correlation matrix format.
The DIFF1/2 value between two supernovae in this preliminary analysis is just defined to be the smallest DIFF1/2 value found
out of all the pairs of spectra with one spectrum drawn from the one supernova and the other from the other supernova.
With no phase restrictions the spectra are drawn from all phases. 
With phase restriction, the spectra are only drawn from the phases allowed by the given restriction. 
(The kind of maximum light used in setting the phase zero point is chosen various ways as discussed in
\S~5.1.) 
The matrix is symmetric, of course.
The diagonal elements are the mean DIFF1/2 values for pairs with supernovae of the same type in the pairs:  
we call these means the self-type means.
The off-diagonal elements are the mean DIFF1/2 values for pairs with supernovae of different types in the pairs:
we call these means the cross-type means.
More digits are shown than are significant to allow for numerical consistency checks and
to give the table a consistent appearance.
}
\end{deluxetable}

\clearpage

\end{document}